\documentclass[a4paper,UKenglish,cleveref, autoref, thm-restate,notab, nosubfigcap,nolineno]{socg-lipics-v2021}
\hideLIPIcs

\usepackage{xcolor}
\usepackage{mathbbol}
\usepackage{tikz}
\usetikzlibrary{matrix,arrows,cd,calc}
\usepackage{thmtools}
\usepackage{thm-restate}

\usepackage{amsfonts}       

\usepackage{algorithm}
\usepackage{algpseudocode}
\usepackage{wrapfig}
\usepackage{graphicx}
\graphicspath{{./figures/}}
\usepackage{amsmath}
\usepackage{amsthm}
\usepackage{amssymb}
\usepackage{mathtools}
\usepackage{enumitem}

\newtheorem{problem}[theorem]{Problem}

\newcommand{\define}[1]{\textit{#1}}

\newcommand{\disk}{\mathsf{disk}}

\newcommand{\field}{\mathbb{k}}
\newcommand{\stief}{\mathsf{V}}

\newcommand{\R}{\mathbb{R}}
\newcommand{\Z}{\mathbb{Z}}
\newcommand{\N}{\mathbb{N}}

\newcommand{\B}{\mathcal{B}}
\newcommand{\X}{\mathcal{X}}

\newcommand{\NNN}{\mathcal{N}}

\newcommand{\UUU}{\mathcal{U}}

\renewcommand{\epsilon}{\varepsilon}

\newcommand{\im}{\mathsf{Im}}
\newcommand{\simp}{\mathsf{simp}}

\newcommand{\fibred}{{\normalfont\textsc{FibeRed}}}

\newcommand{\covandpart}{{\normalfont\textsc{CoverAndPartitionUnity}}}
\newcommand{\nerve}{{\normalfont\textsc{Nerve}}}
\newcommand{\localrep}{{\normalfont\textsc{LocalLinearRepresentation}}}
\newcommand{\estimatefibcoord}{{\normalfont\textsc{EstNormFiberCoordinates}}}
\newcommand{\estimatecocycles}{{\normalfont\textsc{EstCocycles}}}
\newcommand{\alignfibs}{{\normalfont\textsc{AlignFibers}}}
\newcommand{\assemble}{{\normalfont\textsc{Assemble}}}
\newcommand{\estimatetangandnorm}{{\normalfont\textsc{EstTangAndNormBun}}}
\newcommand{\estimatereach}{{\normalfont\textsc{EstReach}}}
\renewcommand{\phi}{\varphi}

\title{FibeRed: Fiberwise Dimensionality Reduction of Topologically Complex Data with Vector Bundles}

\titlerunning{Fiberwise Dimensionality Reduction}

\author{Luis Scoccola}{Department of Mathematics, Northeastern University, USA \and \url{https://luisscoccola.com}}{l.scoccola@northeastern.edu}{https://orcid.org/0000-0002-4862-722X}{supported by the NSF through
grants CCF-2006661 and CAREER award DMS-1943758}

\author{Jose A. Perea}{Department of Mathematics and Khoury College of Computer Sciences, Northeastern University, USA \and \url{https://www.joperea.com/} }{j.pereabenitez@northeastern.edu}{https://orcid.org/0000-0002-6440-5096}{supported by the NSF through grants CCF-2006661 and CAREER award DMS-1943758}

\authorrunning{L. Scoccola and J. A. Perea}

\Copyright{Luis Scoccola and Jose Perea}



\category{}

\supplement{Proof-of-concept implementation \cite{implementation}}

\acknowledgements{The authors thank Matt Piekenbrock for various fruitful conversations.}

\EventEditors{Erin W. Chambers and Joachim Gudmundsson}
\EventNoEds{2}
\EventLongTitle{39th International Symposium on Computational Geometry
(SoCG 2023)}
\EventShortTitle{SoCG 2023}
\EventAcronym{SoCG}
\EventYear{2023}
\EventDate{June 12--15, 2023}
\EventLocation{Dallas, Texas, USA}
\EventLogo{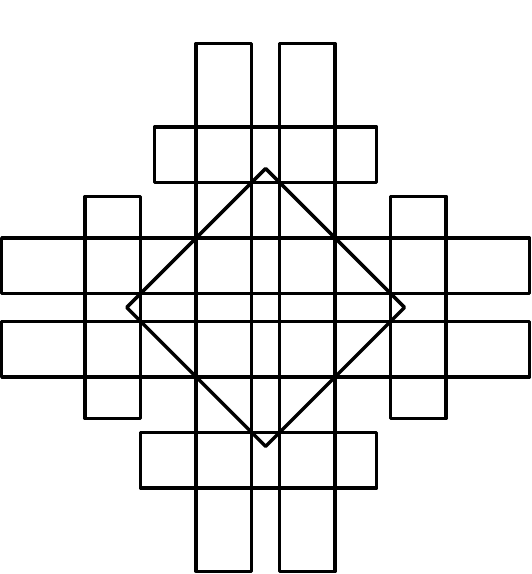}
\SeriesVolume{258}
\ArticleNo{XX}

\begin{document}

\maketitle

\begin{abstract}
Datasets with non-trivial large scale topology can be hard to embed in low-dimensional Euclidean space with existing dimensionality reduction algorithms.
We propose to model topologically complex datasets using vector bundles, in such a way that the base space accounts for the large scale topology, while the fibers account for the local geometry.
This allows one to reduce the dimensionality of the fibers, while preserving the large scale topology.
We formalize this point of view and, as an application, we describe a dimensionality reduction algorithm based on topological inference for vector bundles.
The algorithm takes as input a dataset together with an initial representation in Euclidean space, assumed to recover part of its large scale topology, and outputs a new representation that integrates local representations obtained through local linear dimensionality reduction.
We demonstrate this algorithm on examples coming from dynamical systems and chemistry.
In these examples, our algorithm is able to learn topologically faithful embeddings of the data in lower target dimension than various well known metric-based dimensionality reduction algorithms. 
\end{abstract}

\section{Introduction}

\medskip \noindent\textbf{Motivation.}
We take the manifold hypothesis at face value and consider data consisting of a finite sample of a Riemannian manifold.
We take the goal of \define{dimensionality reduction} to be that of learning an embedding of the input data in low dimension, in such a way that the differentiable structure of the underlying manifold is preserved.
This is different from \define{charting}, whose objective we take to be that of producing local parametrizations of the data that, together, cover the entire manifold.

\begin{figure}
    \begin{center}
    \includegraphics[width=\linewidth]{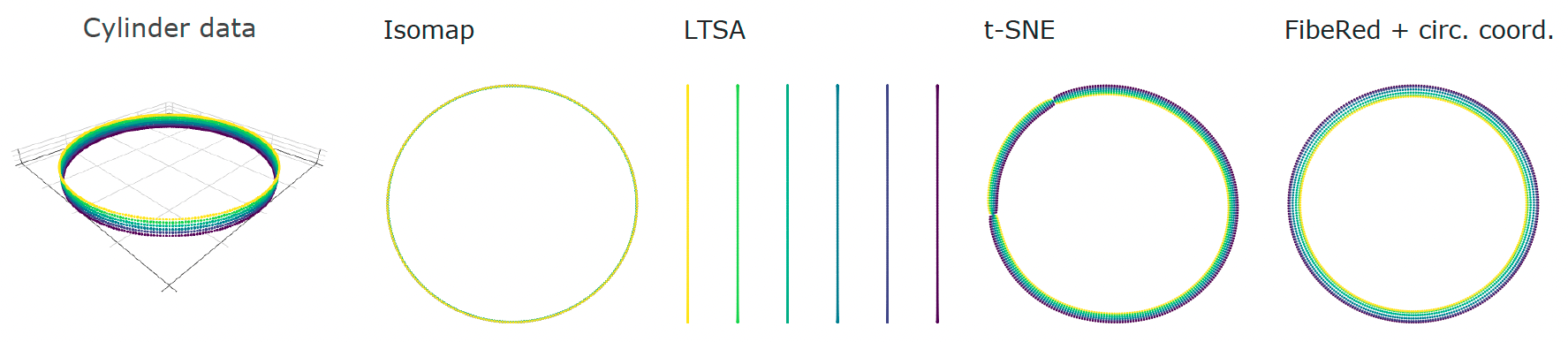}
    \end{center}
    \caption{\textit{Left}: A sample from a cylinder with height equal to $0.15$ times its radius, colored by height.
    The cylinder is developable, since it is diffeomorphic to an annulus in $\R^2$, and is also flat, but it is not isometric to the annulus, which also has a flat, yet distinct, Riemannian metric.
    \textit{Center three}: Well known dimensionality reduction algorithms run on the cylinder data.
    The outputs are representative of other parameter choices and of Laplacian Eigenmaps (LE) \cite{belkin-niyogi}, Diffusion Maps (DM) \cite{coifman-lafon}, LLE \cite{roweis-saul}, HLLE, t-SNE \cite{maaten-hinton}, and UMAP \cite{mcinnes-healy-melville}.
    Some algorithms only capture the circularity, others only the local 2D structure, while others capture both, but they are not able to consistently align the local 2D structure.
    \textit{Right}: The output of fiberwise dimensionality reduction.}
    \label{figure:cylinder-example}
\end{figure}

We refer to dimensionality reduction algorithms which aim to preserve metric relationships and do not explicitly incorporate large scale topology in their objective function as \define{metric-based}.
Metric-based algorithms work best when the Riemannian manifold underlying the data can be isometrically embedded in the target dimension.
For example, algorithms such as Isomap \cite{tenenbaum-silva-langford}, Local Tangent Space Alignment (LTSA) \cite{zhang-zha}, and Hessian Eigenmaps (HLLE) \cite{donoho-grimes}
assume that the manifold $\X$ underlying the data is \define{isometrically developable}, in the sense that 
$\X$ is a $d$-dimensional Riemannian manifold for which there exists an embedded $d$-dimensional manifold $\X' \subseteq \R^d$ and a diffeomorphism $\X' \to \X$ which is a Riemannian isometry.
An isometrically developable manifold $\X$ is necessarily \define{flat} (i.e., locally isometric to Euclidean space) and \define{developable} (i.e., diffeomorphic to an embedded $d$-dimensional manifold $\X' \subseteq \R^d$).
But a manifold can be flat and developable without it being isometrically developable: a simple example is that of a straight cyilinder in $\R^3$ (\cref{figure:cylinder-example}).
As observed in \cite{lee-verleysen}, and shown in \cref{figure:cylinder-example}, already in the setting of a flat and developable $d$-dimensional manifold, metric-based dimensionality reduction algorithms can fail to find an embedding of the data in $\R^d$.
On the mathematical side, while Whitney's embedding theorem \cite{whitney} guarantees that any closed $d$-dimensional manifold admits a smooth ($C^\infty$) embedding in $2d$ dimensions, a smooth, Riemannian isometric embedding of a closed $d$-dimensional Riemannian manifold can require in the order of $d^2$ dimensions \cite{cartan}.
Thus, the preservation of distances requires more complicated embeddings than the preservation of topology.

If we remove a small portion of the cylinder of \cref{figure:cylinder-example}, in order to make it a curved rectangle, most metric-based dimensionality reduction algorithms have no problem finding an embedding in $\R^2$.
It is thus the non-trivial topology of the cylinder---its circularity---that causes difficulties.
This suggests that embeddings of topologically non-trivial manifolds can be built by gluing local representations along a representation of the global topological structure:
in the case of the cylinder, one would try to glue 2D patches around a circle in a globally consistent manner.
This leads to the following problem, formalized as the vector bundle embedding problem (\cref{problem:main-problem}):

\hspace{0.2cm}{\itshape Given a dataset $X$ and an initial map $X \to \R^D$ capturing the large scale topology of $X$, find a new representation $X \to \R^D$ that captures the large scale topology as well as the local geometry.}

We call our approach to the above problem \define{fiberwise dimensionality reduction} (\fibred{}).
In the examples of \cref{section:examples}, we focus on manifolds with an essential loop, and, as initial map, we use circular coordinates based on persistent cohomology \cite{silva-morozov-vejdemo,perea-circular}, a technique from Topological Data Analysis \cite{oudot,ghrist}.
Nevertheless, the approach is not restricted to the case of a circular initial embedding and one could use as initial map one constructed by, e.g., other cohomological coordinates \cite{perea-projective,polanco-perea,scoccola2023}, standard non-linear dimensionality reduction methods \cite{lafon-lee,coifman-lafon}, or lens functions as in \cite[Section~4]{singh-memoli-carlsson}.

\medskip\noindent\textbf{Contributions.}
We show that the theory of vector bundles is useful in abstracting (\cref{section:main-problem}), devising solutions to (\cref{section:main-algorithm}), and computing obstructions to solving (\cref{section:obstructions}) the problem of extending an initial coarse representation of data to a new, more descriptive representation.
We demonstrate with computational examples (\cref{section:examples}) that topological inference for vector bundles can be carried out in practice.
In particular, we show that efficient embeddings and chartings of topologically non-trivial data can be learned with this approach and give examples supporting the claim that metric-based dimensionality reduction algorithms are often not able to find such representations.
We implement our main algorithm in \cite{implementation}.

\medskip\noindent\textbf{Related work.}
Various dimensionality reduction schemes \cite{teh-roweis, roweis-saul-hinton,brand} learn a global alignment of local linear models from the local interactions of the models, which can be challenging in the presence of non-trivial topology.
In contrast, our approach assumes a global topological representation is given and builds and aligns the local linear models along this representation.

There has been recent interest in designing topology-preserving dimensionality reduction schemes \cite{luo-xu-zhang-jin,yan-zhao-rosen-scheidegger-wang,moor-horn-rick-borgwardt,wagner-solomon-bendich}.
Our approach is different from previous approaches we are aware of, as it builds a new representation around an initial topological representation, instead of using topology to regularize an essentially metric objective.

Our cut-unfold technique of
\cref{section:cut-unfold}
has a similar goal to that of \cite{lee-verleysen,yan-zhao-rosen-scheidegger-wang}, which propose to tear a data manifold in order to find efficient representations of it.
A main difference is that our technique allows the user to select a specific hole to cut and to use topological persistence to guide this choice.

\section{The vector bundle embedding problem}
\label{section:main-problem}

For background, please refer to
\cref{section:background}
and the references therein.
In \cref{section:main-problem} we describe the Vector Bundle Embedding problem; in \cref{section:discretization}, we recall the notion of discrete vector bundle that we use to estimate vector bundles from finite samples; and in \cref{section:obstructions} we explain how characteristic classes of vector bundles give computable obstructions to solving the vector bundle embedding problem and can thus be used for parameter selection.

\subsection{Main problem}
\label{section:main-problem}

Let $\B$ be a closed differentiable manifold and let $\pi : \X \to \B$ be a rank $r$ Euclidean vector bundle with zero-section $s_0 : \B \to \X$,
where by \define{Euclidean} we mean that $\pi$ is endowed with a scalar product on each fiber $\pi^{-1}(b) \subseteq \X$, which varies smoothly with $b \in \B$.
The main problem we seek to solve is that of extending an embedding $\B \to \R^D$ to a fiberwise isometric embedding of $\X$, as follows:

\begin{problem}
    \label{problem:main-problem}
Given an embedding $\iota : \B \to \R^D$, find a fiberwise isometric embedding $\overline{\iota} : \X \to \R^D$ that extends $\iota$ in the sense that $\overline{\iota} \circ s_0 = \iota$, and that is orthogonal to $\B$, in the sense that $\overline{\iota}(\pi^{-1}(b))\, \bot \, \iota(T_{b}B)$ for all $b \in \B$.
\end{problem}

By \define{fiberwise isometric embedding} $\X \to \R^D$ we mean a map that is a linear isometry when restricted to each fiber $\pi^{-1}(b) \subseteq \X$, where $b \in \B$.

Let $\nu : N \to \B$ be the normal bundle of the embedding $\iota : \B \to \R^D$ and endow $\nu$ with the Euclidean structure inherited from $\R^D$.
The following result reduces \cref{problem:main-problem} to a problem only involving vector bundles.

\begin{lemma}
    \label{lemma:reduction-main-problem}
    \cref{problem:main-problem} admits a solution if and only if there exists a morphism $\X \to N$ of vector bundles over $\B$ that is an isometry in each fiber.
\end{lemma}

In order to do this, we trivialize the bundles $\X$ and $N$ over a common cover of the base $\B$ and construct the embedding $\X \to N$ by restricting to each element of the cover.
Formally, we proceed as follows.

Let $e$ be the dimension of $\B$, so that the rank of $\nu$ is $D - e$.
Let $\UUU = \{U_i\}$ be a cover of $\B$ such that both $\pi$ and $\nu$ can be trivialized over $\UUU$ and let $\X_i := \pi^{-1}(U_i)$.
Recall that $\stief(n,m)$ denotes the Stiefel manifold, which consist of $m$-by-$n$ matrices with orthonormal columns and that $O(n) = \stief(n,n)$ denotes the orthogonal group.
Let $\alpha = \{\alpha_i : U_i \to \stief(D-e,D)\}$ be local bases for $N$, and let $\Theta = \{\Theta_{ij} : U_i \cap U_j \to O(D-e)\}$ be defined by $\Theta_{ij}(b) = \alpha_i(b) \alpha_j(b)^T$ for all $b \in U_i \cap U_j$, so that $\Theta$ is a cocycle with associated vector bundle $\nu$.
Finally, let $\{(\pi|_{\X_i},f_i) : \X_i \to U_i \times \R^r\}$ be a metric trivialization of $\X$ over $\UUU$ and let $\Omega = \{\Omega_{ij} : U_i \cap U_j \to O(r)\}$ be defined as the unique set of maps satisfying
\begin{equation}
    \label{equation:omega-definition}
    \Omega_{ij}(\pi(x))\,f_j(x) = f_i(x), \text{ for all $x \in \X_i \cap \X_j$},
\end{equation}
so that $\Omega$ is a cocycle with associated vector bundle $\pi$.
We refer to the maps $\{f_i : \X_i \to \R^r\}$ as the \define{fiber coordinates}.
With these definitions, one can use \cref{lemma:reduction-main-problem} to prove the following.

\begin{proposition}
    \label{proposition:cocycle-characterization}
    There exists a fiberwise isometric embedding $\X \to N$ if and only if there exist maps $\Phi = \{\Phi_{i} : U_i \to \stief(r,D-e)\}$ such that
    \begin{equation}
        \label{equation:idealized-objective-function}
        \Phi_i(b) \Omega_{ij}(b) = \Theta_{ij}(b) \Phi_j(b), \; \text{for all $i$ and $j$ and $b \in U_i \cap U_j$}.
    \end{equation}
\end{proposition}

Given the maps $\Phi = \{\Phi_{i} : U_i \to \stief(r,D-e)\}$ of \cref{proposition:cocycle-characterization}, one obtains the fiberwise isometric embedding $\overline{\iota} : \X \to \R^D$ by $\overline{\iota}(x) = \alpha_i(b) \, \Phi_i(b) \, f_i(x) + \iota(b)$, where $b = \pi(x)$.

In general, the fiberwise isometric embedding $\overline{\iota} : \X \to \R^D$ is not an embedding of $\X$ as a manifold, since different fibers may intersect.
Nonetheless, if $\tau > 0$ is the \define{reach} \cite[Definition~2.1]{aamari-kim-chazal-michel-rinaldo-wasserman} of $\iota(\B) \subseteq \R^D$, i.e., the largest possible radius of a uniform tubular neighborhood around $\iota(\B)$, one can find an embedding of a full-dimensional compact subset of $\X$ by scaling the fibers by a fraction of $\tau$, as follows.
Let $\disk(\pi) \subseteq \X$ be the unit disk of the bundle $\pi$, namely, the subspace of points $x \in \X$ such that $\|x - s_0(\pi(x))\| \leq 1$, where $\|-\|$ denotes the norm of the fiber $\pi^{-1}(\pi(x))$ induced by the Euclidean structure of $\pi$.
Then, the following formula gives an embedding $\disk(\pi) \to \R^D$:
\begin{equation}
    \label{equation:assembly-theory}
     x \;\mapsto\; c\,\tau \cdot \alpha_i(\pi(x)) \, \Phi_i(\pi(x)) \, f_i(x) + \iota(\pi(x)), \;\text{ for $\pi(x) \in U_i$,}
\end{equation}
where $0 < c < 1$ is any fixed constant.

\subsection{Vector bundles from finite samples}
\label{section:discretization}

In practice, continuous maps to a Stiefel manifold or orthogonal group---such as the maps $\{\alpha_i : U_i \to \stief(D-e,D)\}$ or the cocycle $\{\Omega_{ij} : U_i \cap U_j \to O(r)\}$ of \cref{section:main-problem}---are hard to work with, as they are potentially determined by an infinite amount of data.
One of the main takeaways of \cite{scoccola-perea} is that one can work with Euclidean vector bundles in practice by considering only constant maps into Stiefel manifolds or orthogonal groups.
In order to accomplish this, one relaxes the notion of Euclidean vector bundle as follows.

Given a simplicial complex $S$, a rank $r$ \define{discrete approximate} cocycle on $S$ (\cite[Definition~5.1]{scoccola-perea}) consists of a family of matrices $\{\Omega_{ij} \in O(r)\}$ indexed by the oriented $1$-simplices of $S$, which satisfies $\Omega_{ij} = \Omega_{ji}^T$.
There is a similar way of discretizing maps into a Stiefel manifold (\cite[Definition~5.4]{scoccola-perea}).
These discretizations can be used to represent usual vector bundles \cite[Theorem~A]{scoccola-perea} and any vector bundle can be represented in this way \cite[Proposition~5.7]{scoccola-perea}.

This justifies the fact that, in \cref{section:main-algorithm}, we discretize the base $\B$ by considering the simplicial complex given by the nerve of a cover $\UUU = \{U_i\}$ and we consider constant maps from $U_i$ into a Stiefel manifold and from $U_i \cap U_j$ into an orthogonal group.

\subsection{Computable obstructions to vector bundle embedding}
\label{section:obstructions}

The theory of vector bundles provides us with algebraic obstructions to solving \cref{problem:main-problem}, namely, characteristic classes.
We now give a few details about the subject;
we refer the reader to \cite{milnor-stasheff} for a detailed account of the theory of characteristic classes.

To a vector bundle $\pi : \X \to \B$ and number $i \in \N$, one can associate an element $w_i(\pi) \in H^i(\B;\Z/2)$ of the $i$th cohomology group of $\B$ with coefficients in the group $\Z/2$, called the \define{$i$th Stiefel--Whitney class} of $\pi$.
This procedure is such that, if $\pi$ and $\pi'$ are isomorphic vector bundles over the same base $\B$, then $w_i(\pi) = w_i(\pi')$.

If \cref{problem:main-problem} admits a solution, then there exists a complement of $\pi$ in $\nu$, that is there exists a vector bundle $\kappa$ over $\B$ such that $\pi \oplus \kappa \cong \nu$, where $\oplus$ denotes the direct sum of vector bundles.
It follows from the Whitney product formula \cite[Section~4,~Axiom~3]{milnor-stasheff} that $w(\pi) \smallsmile w(\kappa) = w(\nu)$, where $\smallsmile$ denotes the cup-product in cohomology \cite[Section~3.2]{hatcher}.
In particular, when \cref{problem:main-problem} admits a solution, we have the following:
\begin{itemize}[leftmargin=8mm]
    \item If $D = r + e$, then $w_1(\pi) = w_1(\nu) \in H^1(\B; \Z/2)$.
    \item If $D = r+e+1$, then $w_2(\pi) - w_1(\pi)^2 + w_1(\pi) \smallsmile w_1(\nu) = w_2(\nu) \in H^2(\B; \Z/2)$.
\end{itemize}

Thus, if any of these equalities is not satisfied, then \cref{problem:main-problem} does not admit a solution.
These obstructions can be computed from finite samples using \cite[Theorem~C]{scoccola-perea}.

\section{The fiberwise dimensionality reduction scheme}
\label{section:main-algorithm}

We describe the \fibred{} algorithm in \cref{section:main-routine,section:subroutines,section:minimizing-objective,section:preprocessing}. 
In \cref{section:justification-fiber-coordinates} we justify a main subroutine of the algorithm.
In \cref{section:choosing-parameters}, we explain how we choose parameters.
We represent vector bundles using discrete approximate cocycles as in \cite{scoccola-perea} (see \cref{section:discretization}).

\begin{figure}
    \begin{center}
        \scalebox{0.9}{
        {\def\svgwidth{1\textwidth}
        \footnotesize
        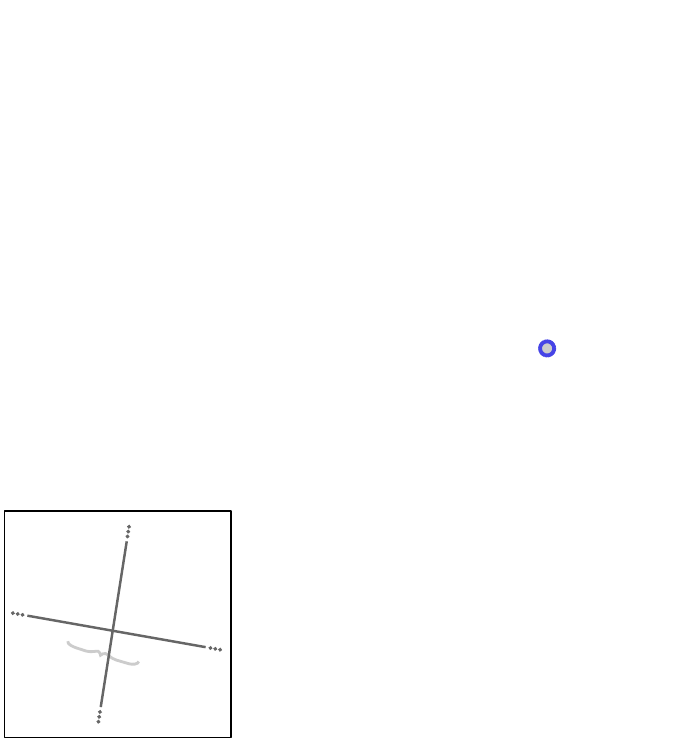}}
    \end{center}
    \caption{Schematic representation of the main constructions in the \fibred{} pipeline.}
    \label{figure:schematic-description-algorithm}
\end{figure}

To facilitate the interpretation of the different steps of the algorithm, the notation is kept as in \cref{section:main-problem}, except for the spaces $\X$ and $\B$, which we denote here by $X$ and $B$ to emphasize the fact that we are working with finite samples $X \subseteq \X$ and $B \subseteq \B$.
See also \cref{figure:schematic-description-algorithm} for a schematic representation of some of the main steps (4,5,6) of the algorithm.

Precise assumptions about the input of the algorithm are in
\cref{section:assumptions}.
We remark that our algorithm can be efficiently implemented; we give more details in
\cref{section:efficiency}.

\subsection{Main routine}
\label{section:main-routine}

\begin{wrapfigure}{R}{0.6\textwidth}
\begin{minipage}{0.6\textwidth}
\vspace{-0.8cm}
\begin{algorithm}[H]
\begin{algorithmic}[1]
\State $\UUU, \rho \gets \covandpart(k,B)$
\State $\NNN \gets \nerve(B,\UUU)$
\For{$1 \leq i \leq k$}
    \State $\ell_i \gets \localrep(X,\UUU,d,i)$
    \State $\Psi_i, \alpha_i \gets \estimatetangandnorm(B,\UUU,e,i)$
    \State $\overline{f}_i \gets \estimatefibcoord(B,\Psi_i,\ell_i)$
\EndFor
\State $\tau \gets \estimatereach(B,\UUU,\Psi)$
\For{$(ij) \in \NNN$}
    \State $\Omega_{ij},\Theta_{ij} \gets \estimatecocycles(\overline{f}_i,\overline{f}_j,\alpha_i,\alpha_j)$
\EndFor
\State $\Phi \gets \alignfibs(\NNN, \Omega, \Theta, \texttt{n\_iter})$
\State \Return $\assemble(\rho, \tau, c, \alpha, \Phi, \overline{f}, \pi)$
\end{algorithmic}
    \caption{$\fibred(X,\pi,k,e,d,c,\texttt{n\_iter})$}
    \label{alg:cap}
\end{algorithm}
\vspace{-1.4cm}
\end{minipage}
\end{wrapfigure}

\textbf{Inputs.}
A dataset represented by a finite set $X$ together with a distance matrix $\partial : X \times X \to \R$; and a function $\pi : X \to \R^D$.
We let $B := \pi(X) \subseteq \R^D$.
 
\noindent\textbf{Parameters.}
A number $k \in \N$, the number of sets we use to construct a cover of $B$;
a number $\texttt{n\_iter} \in \N$ used in the $\alignfibs$ subroutine;
an estimate $e \in \N$ of the intrinsic dimension of $\B$; an estimate $d \in \N$ of the intrinsic dimension of $\X$; a fiber scale $0 < c < 1$.

\noindent\textbf{Output.}
A map $X \to \R^D$.

\medskip

With this notation, the map $\iota$ of \cref{section:main-problem} corresponds to the inclusion $B = \pi(X) \subseteq \R^D$ and the rescaling of $\overline{\iota}$ of \cref{equation:assembly-theory} corresponds to the output of the algorithm.

\subsection{Subroutines}
\label{section:subroutines}

\noindent\textbf{Compute cover and partition of unity (\covandpart).}
We compute a cover $\UUU = \{U_i \subseteq B\}_{1 \leq i \leq k}$ of $B$ as follows.
We first run on $B$ an approximate algorithm for the $k$-center problem.
We use a simple, greedy approach, but more sophisticated options are available (see, e.g., \cite{garcia-menchaca-menchaca-pomares-perez-lakouari} for a survey).
This results in $k$ points $\{b_1, \dots, b_k\} \subseteq B$ and in a radius $c > 0$ such that any point of $B$ is at distance at most $c$ from some $b_i$.
We then let $U_i = \{b \in B : \|b - b_i\| < 3c\}$.
The factor of $3$ is arbitrary; we choose it to ensure that elements of the cover have sufficiently large intersections.

We compute a partition of unity $\rho = \{\rho_i : U_i \to \R\}$ subordinate to $\UUU$ by first defining $p_i(x) = \exp\left( -1/(1 - (\|x - b_i\|/(3c))^2) \right)$ for $x \in U_i$ and $p_i(x) = 0$ for $x \not\in U_i$, and then normalizing as follows $\rho_i(x) = p_i(x) / \sum_j p_j(x)$.

\medskip \noindent\textbf{Compute nerve of cover (\nerve).}
We let $\NNN$ be the undirected graph with vertices $1 \leq i \leq k$
and an edge $(ij)$ with weight $s_{ij} = |U_i \cap U_j|$ when $U_i \cap U_j \neq \emptyset$.

\medskip \noindent\textbf{Compute local linear representation (\localrep).}
Given $1 \leq i \leq k$, we let $X_i := \pi^{-1}(U_i)$ and apply a linear dimensionality reduction algorithm to each $X_i$, resulting in a function $\ell'_i : X_i \to \R^d$.
In our implementation, we use classical multidimensional scaling (see, e.g., \cite{borg-groenen}).
We then mean-center $\ell'_i$ to get a function $\ell_i : X_i \to \R^d$.

\medskip \noindent\textbf{Estimate local trivialization of tangent and normal bundle (\estimatetangandnorm).}
Given $1 \leq i \leq k$,
we compute an orthonormal frame $\Psi_i \in \stief(e,D)$ by applying PCA with target dimension $e$ to $U_i \subseteq \R^D$.
We then compute an orthonormal frame $\alpha_i \in \stief(D-e,D)$ such that $\alpha_i \perp \Psi_i$.

\medskip \noindent\textbf{Estimate normalized fiber coordinates (\estimatefibcoord).}
Given $1 \leq i \leq k$, we define $t : X_i \to \R^{e}$ by $t(x) = \Psi_i^T (\pi(x) - b_i)$.
We find a linear transformation $m_i : \R^{d} \to \R^{e}$, which has minimal Frobenius norm and minimizes
\begin{equation}
    \label{equation:fiber-coordinate-fit}
    \sum_{x \in X_i} \|t(x) - m_i(\ell_i(x))\|^2,
\end{equation}
and compute an orthonormal frame $\eta_i \in \stief(r,d)$ with image in the kernel of $m_i$.
We let $f_i := \eta_i^T \circ \ell_i : X_i \to \R^{r}$, and obtain a normalized fiber coordinate $\overline{f}_i : X_i \to \R^{r}$ with image contained in the unit ball by normalizing $f_i$.
We justify these choices in \cref{section:justification-fiber-coordinates}.

\medskip \noindent\textbf{Estimate reach (\estimatereach).}
If $\{b_1, \dots, b_k\} \subseteq B$ are the centers of the $k$ balls used to construct the cover $\UUU$ in \covandpart, we compute an estimate of the reach of $B$ by
\[
    \tau = \inf_{i\neq j} \frac{\|b_j - b_i\|^2}{2 \sqrt{\|b_j - b_i\|^2 - \|\Psi_i^T (b_j - b_i)\|^2} }.
\]
This formula is equivalent to \cite[Equation~6.1]{aamari-kim-chazal-michel-rinaldo-wasserman}, where it is proven that, under suitable assumption, it yields a consistent estimator of the reach.

\medskip \noindent\textbf{Estimate cocycles for $\nu$ and $\pi$ (\estimatecocycles).}
Based on \cref{equation:omega-definition}, given $(ij) \in \NNN$, we compute an orthogonal matrix $\Omega_{ij} \in O(r)$ which minimizes 
\[
    \sum_{x \in X_i \cap X_j} \|\Omega_{ij} f_j(x) - f_i(x)\|^2.
\]
We also compute an orthogonal matrix $\Theta_{ij} \in O(D - e)$ which minimizes $\|\Theta_{ij} - \alpha_i^T \alpha_j\|_F$, where $\|-\|_F$ denotes the Frobenius norm.
Both minimizations are instances of the orthogonal Procrustes problem, which can be solved using SVD (see, e.g., \cite[Section~7.4]{horn-johnson}).

\medskip \noindent\textbf{Align fibers (\alignfibs).}
Based on \cref{equation:idealized-objective-function},
we compute orthonormal frames $\{\Phi_i \in \stief(r,D - e)\}$ minimizing the following expression; we describe the minimization procedure in \cref{section:minimizing-objective}:
    \begin{equation}
        \label{equation:objective-function}
        \sum_{(ij) \in \NNN} s_{ij} \; \| \Phi_i \Omega_{ij} -  \Theta_{ij} \Phi_j\|_F.
    \end{equation}

\medskip \noindent\textbf{Compute final representation (\assemble).}
Based on Eq.~\ref{equation:assembly-theory}, we represent $x \in X$ by
    \[
        \sum_{1 \leq i \leq k} \rho_i(x) \;\big( c\tau \cdot \alpha_i\, \Phi_i\, \overline{f}_i(x) + \pi(x) \big).
    \]

\subsection{Minimizing \cref{equation:objective-function}}
\label{section:minimizing-objective}

The minimization problem in \alignfibs{} is non-convex, so a possible solution is to do gradient descent in a product Stiefel manifold. 
This is the approach we take, except that we avoid explicitly computing a gradient, and take a sampling based approach, as done in, e.g., LargeVis \cite{tang-liu-zhang-mei}.
Before describing the approach, we note that, in the case $D = r + e$, the Stiefel manifold $\stief(r,r)$ is equal to the orthogonal group $O(r)$, which is disconnected.
Thus, in this case, any local optimization approach to minimizing \cref{equation:objective-function}, such a gradient descent, is bound to fail.
In \cref{section:preprocessing} we describe a procedure based on the notion of synchronization (see, e.g., \cite{singer}) that reduces the problem from having to align using matrices in $O(r)$ to using matrices in $SO(r)$, which is connected.

\medskip \noindent\textbf{Iterative procedure.}
We start by initializing $\{\Phi_i \in \stief(r,D-e)\}$ at random and setting $a = 1$.
For $1 \leq n \leq \texttt{n\_iter}$, we proceed as follows.
We sample an edge $(ij) \in \NNN$ with probability proportional to its weight $s_{ij}$, let $M$ be an orthonormal frame minimizing $\|M \Omega_{ij} -  \Theta_{ij} \Phi_j\|_F$, and replace $\Phi_i$ with a closest orthonormal frame to the convex combination $(1-a) \Phi_i + a M$.
Finally, we replace $a$ with $1 - n/\texttt{n\_iter}$.

\subsection{Preprocessing in the case $D = r+e$}
\label{section:preprocessing}
In this case, the matrices $\Phi$, $\Omega$, and $\Theta$ are in $O(r)$.
The preprocessing consists of replacing the matrices $\{\Theta_{ij}\}$ by matrices that induce an equivalent problem to the one of minimizing \cref{equation:objective-function}, but for which the matrices $\{\Phi_i\}$ we look for can be taken to be in the special orthogonal group $SO(r)$, which is connected.

Note that, if we want $\Phi_i \Omega_{ij}$ and $\Theta_{ij} \Phi_j$ to belong to the same connected component of $O(r)$, then we must have $\det(\Omega_{ij})\det(\Theta_{ij}) = \det(\Phi_i)\det(\Phi_j) \in O(1) = \{-1,+1\}$.
This suggests that we can let $\omega_{ij} = \det(\Omega_{ij})\det(\Theta_{ij}) \in O(1)$ and consider first the problem of finding $\{\lambda_i \in O(1)\}$ such that $\lambda_i \lambda_j = \omega_{ij}$, which leads to minimizing the objective function
\[
        \sum_{(ij) \in \NNN} s_{ij} \; |\omega_{ij} - \lambda_i \lambda_j|^2.
\]
This is a well known synchronization problem, for which an approximate solution can be found effectively and efficiently with spectral methods \cite{singer-wu,bandeira-singer-spielman}.
Here, we use \cite[Algorithm~2.3]{bandeira-singer-spielman}, with $d = 1$, which yields an approximate solution $\{\lambda_i \in O(1)\}$.

Given $\lambda \in O(1) = \{-1,+1\}$ let $M(\lambda) \in O(r)$ be the diagonal matrix with all diagonal entries equal to $1$, except for the first one, which is equal to $\lambda$.
With this in mind, we can replace $\Theta_{ij}$ by $M(\lambda_i) \Theta_{ij} M(\lambda_j)$.
Having done this, we can now restrict the matrices $\{\Phi_i\}$ to belong to the connected component of $O(r)$ of orthogonal matrices with $+1$ as determinant.
More specifically, we now can carry out the optimization procedure described above, but restring the matrices $\{\Phi_i\}$ to be in $SO(r) \subseteq O(r) = \stief(r,r)$.
 
\subsection{Justification of estimate of fiber coordinates}
\label{section:justification-fiber-coordinates}

Let $x_i := s_0(b_i) \in \X$.
We interpret the local model $\ell_i : X_i \to \R^d$ as a projection $\ell_i : X_i \to T_{x} \X \cong \R^d$ of $X_i$ onto the tangent space at the origin of the fiber $\pi^{-1}(b_i)$.
In the idealized case
(\cref{section:assumptions}),
the fiber coordinate $f_i : \X_i \to \R^r$ is given by any map fitting into a fiberwise isometric diffeomorphism $(\pi|_{\X_i}, f_i) : \X_i \to U_i \times \R^r$.
When dealing with finite samples, we use the following composite as a proxy for $f_i$:
\[
    X_i \xrightarrow{\ell_i} T_{x_i} \X \xrightarrow{(d f_i)_{x_i}} T_{f_i(x_i)}\R^r \cong \R^r.
\]
Note that, by assumption, $(d f_i)_{x_i}$ is the second component of an isometric isomorphism of Euclidean vector spaces $d(\pi, f_i)_{x_i} : T_{x_i} \X \to T_{b_i} \B \oplus \R^r$, in which the two direct summands are orthogonal.
It is thus sufficient to estimate the first component $d \pi_{x_i} : T_{x_i} \X \to T_{b_i} \B$ and to then compose $\ell_i$ with the orthogonal projection onto the orthogonal complement of $d\pi_{x_i}$.
We do have an estimate for the composite
\[
    X_i \xrightarrow{\ell_i} T_{x_i} \X \xrightarrow{d \pi_{x_i}} T_{b_i}\B \cong \R^e,
\]
namely $t = \Psi_i^T \circ \pi|_{X_i} : X_i \to \R^{e}$, but, since the embedding $\B \subseteq \R^D$ is not required to preserve the Riemannian structure of $\B$ inherited from that of $\X$, the map $t$ is an approximation of $d\pi_{x_i} \circ \ell_i$ up to a linear map $m_i : \R^d \to \R^e$.
This justifies finding $m_i$ by minimizing \cref{equation:fiber-coordinate-fit},
and getting the approximate fiber coordinate $f_i$ by composing $\ell_i$ with the orthogonal projection onto the kernel of $m_i$.

\subsection{Choosing input and parameters}
\label{section:choosing-parameters}

We discuss some guiding principles to choose parameters for our pipeline.
We focus mostly on parameter selection for the examples of \cref{section:examples}.

\medskip \noindent\textbf{Parameters.}
An estimate of the dimensions $e$ of $B$ and $d$ of $X$ can be obtained by analyzing the explained variance of PCA applied to each of the sets $U_i$ and $X_i$ with a range of target dimensions, but more sophisticated algorithms are available; see, e.g., \cite{little-lee-jung-maggioni}.
The parameter $k$ is chosen to be large enough so that the cover $\UUU$ captures the topology of the base space $B$, and such that each open ball of the cover is sufficiently small so that it can be approximated reasonable well by a linear space.
Admittedly, this is in general a difficult choice and producing good covers of data is an interesting problem in its own right.
In our case, when the base space is the circle, we use $k = 16$; see also \cref{section:parameter-sensitivity} for a parameter sensitivity analysis.
The algorithm is robust to the choice of parameter $\texttt{n\_iter}$, which we choose to be $1000$ in all of our examples.

\medskip \noindent\textbf{Choosing base map and $D$.}
We construct the initial map $\pi : X \to \R^D$ in two ways.

The first way is to use the persistent cohomology of the initial data $X$ to construct circular coordinates $X \to S^1$ and then embed the circle $S^1$ as the unit circle in the plane spanned by the first two coordinates of $\R^D$, $D \geq 2$, which gives us the initial map $\X \to \R^D$.
In order to choose the embedding dimension $D$, we compute the Stiefel--Whitney obstructions, as in \cref{section:obstructions}.
Since the base space is the circle, which is $1$-dimensional, only the first Stiefel--Whitney class provides an obstruction.
The Stiefel--Whitney class of the normal bundle of the embedding $S^1 \subseteq \R^D$ is trivial.
Thus, if the first Stiefel--Whitney class of the estimated cocycle $\Omega$ is trivial, we set $D = r + 1$, and if it is non-trivial, we set $D = r + 2$.

The second way is to use the cut-unfold technique, explained in
\cref{section:cut-unfold},
with the circular coordinates and map $\X \to  \R^D$ by embedding the interval $[0,1)$ as the unit interval of the line spanned by the first coordinate of $\R^D$.
In this case, since the interval is topologically trivial (contractible), the Stiefel--Whitney classes give no obstructions, and thus we set $D = r + 1$.

\section{Examples}
\label{section:examples}

We apply \fibred{} to three examples.
We reproduce a dynamical system simulation from~\cite{charo-artana-sciamarella} and reconstruct an attractor---a torus.
We reconstruct the conformation space---a M\"obius band---and energy landscape of the pentane molecule from a simulation using \texttt{RDKit} \cite{rdkit}; this is inspired by an analysis in \cite{membrillo-pirashvili-steinberg-brodzki-frey}.
Finally, we reconstruct the conformation space of the cyclooctane molecule---a Klein bottle glued to a $2$-sphere---using the data of \cite{martin-thompson-coutsias-watson}.

We compare \fibred{} to various well known dimensionality reduction algorithms (see
\cref{section:other-runs}
for more results).
Given that we consider topologically non-trivial data, we follow \cite{rieck-leitte,paul-chalup} and evaluate the output of algorithms using persistent homology and persistence diagrams (PDs) to quantify the preservation of large scale topology (see
\cref{section:background}
for background and references).
When we do not clarify the field of coefficients used to compute a PD, the PD is independent of this choice.
See \cref{table:minimal-target-dimensions} for a summary.

\begin{figure}
    \begin{center}
    \includegraphics[width=0.8\linewidth]{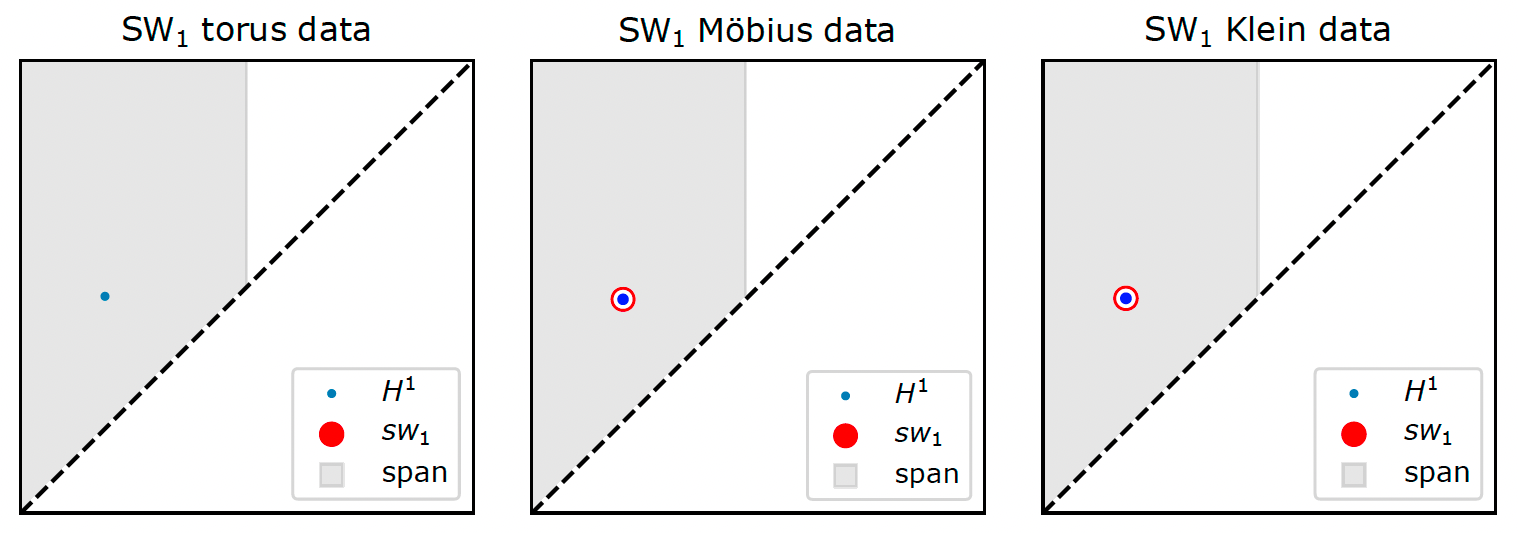}
    \end{center}
    \caption{
    We proceed as in
    \cref{section:obstructions-examples}.
    The persistence diagram of $\{\NNN_r\}_{r \in [0,1]}$ for each of the three examples, with the span of $\Omega$ shaded in grey, and the classes summing to $w_1(\Omega)$ circled in red.
    Recall that in the three examples, the nerve $\NNN$ is a circle and thus the persistence diagram consists of just one prominent 1-dimensional cohomology class.
    In the case of the torus, the first Stiefel--Whitney class is zero and thus there is no obstruction to choosing $D = 1 + 2$ ($1$ being the dimension of the circle and $2$ the rank of the vector bundle).
    In the case of the M\"obius band, the first Stiefel--Whitney class coincides with the only point in the persistence diagram and is thus non-trivial, which gives an obstruction to selecting $D = 1 + 1$, which reflects the fact that the M\"obius band cannot be embedded in the plane.
    Similarly, in the case of the Klein bottle, the Stiefel--Whitney computation gives an obstruction to selecting $D = 1 + 2$, which reflects the fact that the Klein bottle cannot be embedded in $\R^3$.}
    \label{figure:sw-obstructions}
\end{figure}

For the initial map $\pi : X \to B$ we use the implementation of circular coordinates in \cite{dreimac}.
The parameters for \fibred{} are chosen as in \cref{section:choosing-parameters} and the computed Stiefel--Whitney obstructions are in \cref{figure:sw-obstructions}.
For persistent homology computations, we use \texttt{ripser} \cite{bauer} on geodesic distance, estimated as shortest path distance in a 15-nearest neighbor graph.
For other dimensionality reduction algorithms, we use their \texttt{scikit-learn} implementation \cite{scikit-learn}.

An implementation and Jupyter notebooks to reproduce the examples is in \cite{implementation}.

\begin{table}[ht]
    \footnotesize
    \begin{center}
    \begin{tabular}{c|c|c|c|c|c|c|c|c|c|}
        \cline{2-10}
        \rule{0pt}{2ex}    
        & Optimal & Isomap & t-SNE & LE/DM      & LLE & HLLE & LTSA & UMAP       & FibeRed     \\ \hline
\multicolumn{1}{|l|}{\rule{0pt}{2ex}    Cyl.}     & 2              & 3      & 3     & 3          & 3   & 3    & 3    & 3          & \textbf{2} \\ \hline
\multicolumn{1}{|l|}{\rule{0pt}{2ex}Torus}        & 3              & 4      & 4     & 4          & 4   & 4    & 4    & 4          & \textbf{3} \\ \hline
\multicolumn{1}{|l|}{\rule{0pt}{2ex}M\"{o}b.}  & 3              & 4      & 4     & $\mathbf{3^\ast}$ & N/A & N/A  & N/A  & $\mathbf{3^\ast}$ & \textbf{3} \\ \hline
\multicolumn{1}{|l|}{\rule{0pt}{2ex}Klein} & 4              & 5      & 5     & 7          & 7   & 5    & 5    & \textbf{4} & \textbf{4} \\ \hline
\end{tabular}
    \end{center}
    \caption{The minimal target dimension that can be chosen for each of the algorithms considered in this section, so that there exist parameters that return a topologically faithful embedding of the data.
    ``Optimal'' refers to the theoretical minimal embedding dimension.
    ``Cylinder'' refers to the dataset of \cref{figure:cylinder-example}.
    ``Torus'', ``M\"obius band'', and ``Klein bottle'' refer to the three datasets considered in this section.
    Since the M\"obius band data is not Euclidean, some algorithms cannot be run on these data; we denote this with ``N/A''.
    Asteriscs indicate that the data had to be preprocessed with MDS and target dimension $20$ in order to get a topologically faithful embedding with the corresponding algorithm and dimension.
    }
    \label{table:minimal-target-dimensions}
\end{table}

\medskip\noindent\textbf{Torus from attractor of double-gyre dynamical system.}
Dynamical systems can be analyzed by studying the topology of their attractors \cite{abarbanel,perea}.
Given a real-valued time series coming from measurements of a given particle on which a dynamical system acts, one can obtain a pointcloud by constructing a delay embedding of the time series, which, under certain conditions, is concentrated around a diffeomorphic copy of the attractor the particle is converging to \cite{perea}.
Using the delay embedding method with target dimension $4$, it was shown in \cite[Section~4.1]{charo-artana-sciamarella} that a certain attractor of the double-gyre dynamical system \cite{shadden-lekien-marsden} is orientable and has the homology of a torus.
Here, we reproduce the simulation of \cite{charo-artana-sciamarella} using the code from \cite{fernandez} and apply dimensionality reduction to this $4$D pointcloud, with the goal of embedding the attractor and its dynamics in $\R^3$.

\begin{figure}
    \begin{center}
    \includegraphics[width=\linewidth]{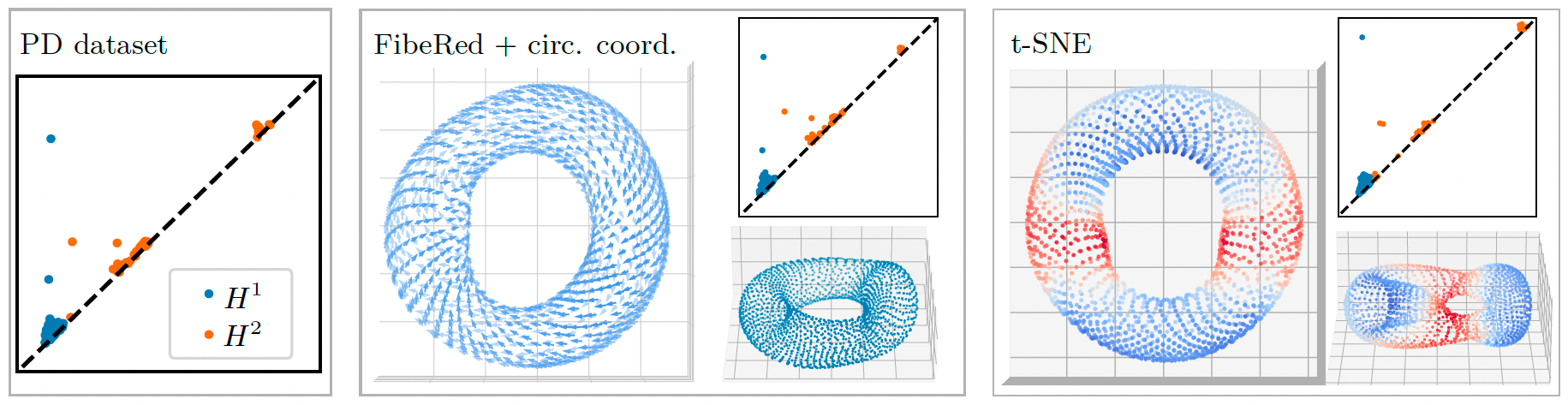}
    \end{center}
    \caption{The PD of the original pointcloud (two prominent $1$-dimensional classes, and one prominent $2$-dimensional class);
    the output of \fibred{} with the reconstructed dynamics and side view, and the PD of the output (which matches the PD of the original pointcloud well);
    the output of t-SNE on the same data and side view, colored by $\kappa$ (red is smaller), there seem to be two self-intersections, and the PD of the output of t-SNE, which has one prominent $1$-dimensional hole and two $2$-dimensional voids, confirming that the red regions have been pinched.
    }
    \label{figure:torus-example}
\end{figure}

In \cref{figure:torus-example}, we show the results of \fibred{} and t-SNE.
In order to highlight self-intersections in low-dimensional representations, we use the following function: given a dataset $X$ and a representation of it $f : X \to Y$ let $\kappa : X \to \R$ be defined by $\kappa(x) = \min_{y \in X} d_Y(f(x),f(y))/d_X(x,y)$.
The output of t-SNE in \cref{figure:torus-example} is representative of the output with other parameter choices and other dimensionality reduction algorithms we have tried on this data (LE, DM, LLE, HLLE, Isomap, UMAP): if the target dimension is 3, there are always self-intersections or tears.
The difficulty faced by metric-based algorithms in this example is that the input torus in 4D has an approximately flat metric and thus it does not admit a smooth isometric embedding in $\R^3$.

\medskip\noindent\textbf{M\"obius band from conformation space of pentane.}
Any fixed molecule admits different realizations, or \define{conformations}, in three-dimensional space.
In, e.g., molecular dynamics \cite{haile}, one is interested in understanding all possible conformations of a molecule.
The collection of conformations up to rotations and translations is known as the \define{conformation space} of the molecule.
Each conformation has an associated energy and the conformation space together with the energy function is known as the \define{energy landscape} of the molecule.

\begin{figure}
    \begin{center}
        \includegraphics[width=\linewidth]{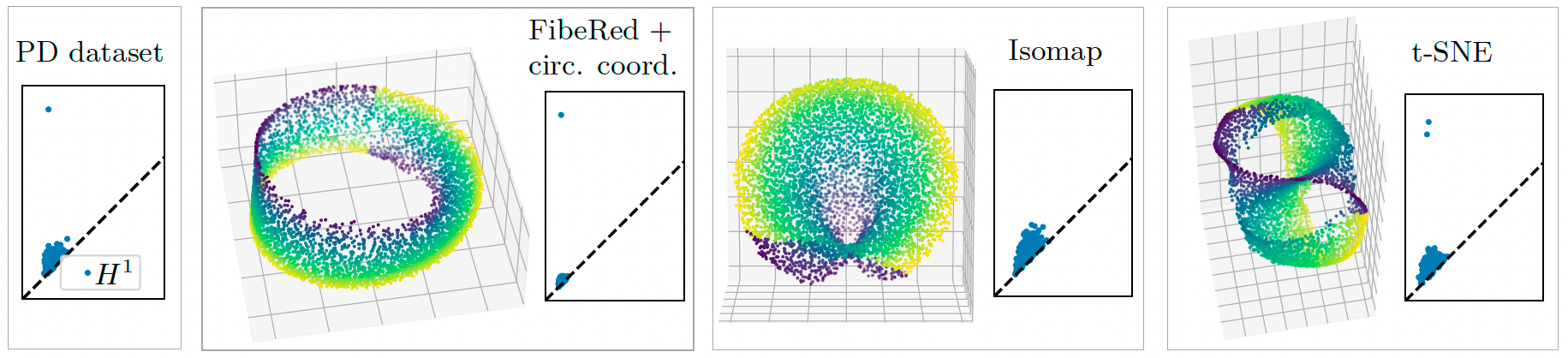}
    \end{center}
    \caption{The PD of the original pointcloud, which has one prominent 1-dimensional class;
    the output of \fibred{} and its PD;
    the output of Isomap and its PD
    (regardless of the parameter for Isomap, the algorithm is unable to capture the circularity of the data, and thus its PD has no prominent features);
    the output of t-SNE and its PD
    (regardless of the parameters for t-SNE, the algorithm is unable to capture the circularity and non-orientability of the data without tears, which cause the output to have two holes).
    Outputs are colored by the (aligned) fiber coordinates estimated by \fibred{}.
    }
    \label{figure:mobius-example}
\end{figure}

\begin{figure}
    \begin{center}
        \includegraphics[width=\linewidth]{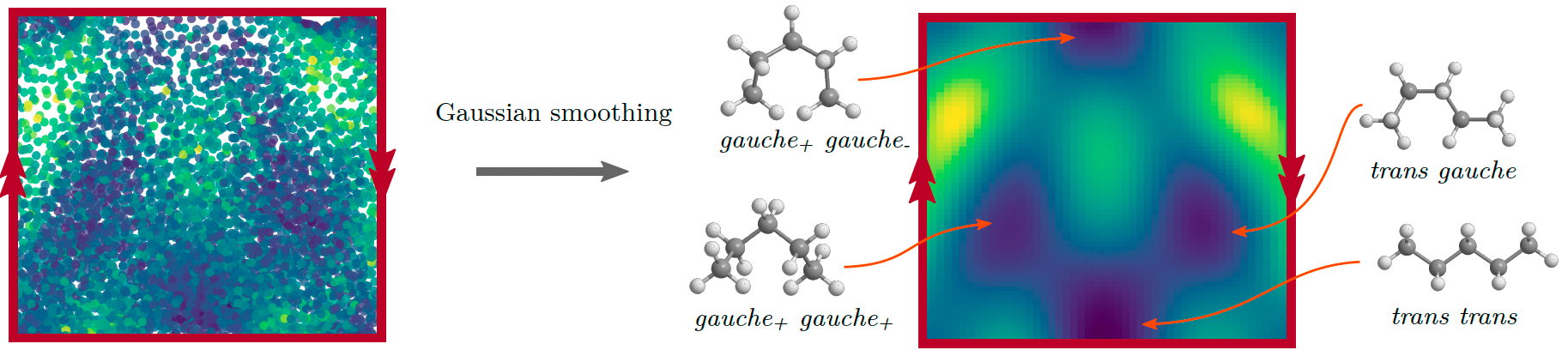}
    \end{center}
    \caption{\textit{Left}: the 2D representation of the conformation space of pentane using \fibred{} with the cut-unfold technique, colored by energy.
    Arrows indicate how the data must be glued in order to recover its global topology;
    this information can be extracted from the cocycle $\Omega$ of $\estimatecocycles$.
    \textit{Right}: a 2D representation of the energy landscape of pentane, where the energy is estimated using the representation on the left and Gaussian smoothing.
    We see that there are four local minima of the energy function.
    By going back to the molecule simulation, we confirm that these four minima correspond to the four well known conformations of pentane \cite{balabin}.}
    \label{figure:mobius-example-conformation}
\end{figure}

We reconstruct the conformation space and energy landscape of the pentane molecule from a simulation (see
\cref{section:pentane-data-generation}
for details).
The pentane molecule has two rotational degrees of freedom (modelled as a torus $S^1 \times S^1$) but also has a symmetry which interchanges the two angles of rotation.
For this reason, the (unlabeled) conformation space consists of a quotient of the torus, which can be seen to be a M\"obius band.
In \cref{figure:mobius-example}, we embed the conformation space of pentane in $\R^3$ and compare the output of \fibred{} to that of Isomap and t-SNE.
LE and DM are able to recover a M\"obius band in $\R^3$; since UMAP uses LE as initialization, it is also able to recover the M\"obius band in $\R^3$.
In \cref{figure:mobius-example-conformation}, we use the cut-unfold technique to find a fundamental domain of the conformation space and estimate the energy landscape.

The difficulty faced by some of the metric-based dimensionality reduction algorithms in this example is that, with respect to the intrinsic metric, the ratio between the height of the M\"obius band and its circumference is approximately $2/3$ and thus there is no isometric embedding in $\R^3$ \cite[Theorem~15.1]{halpern-weaver}.

\medskip\noindent\textbf{Klein bottle from conformation space of cyclooctane.}
In this example, we reconstruct the conformation space and energy landscape of the cyclooctane molecule using the dataset of \cite{martin-thompson-coutsias-watson}.
In \cite{martin-thompson-coutsias-watson}, it is shown that the conformation space of cyclooctane consists of a 2-sphere glued to a Klein bottle along two disjoint circles
and a parametrization of the dataset is given using Isomap and knowledge about how the data was generated.

\begin{figure}
    \begin{center}
        \includegraphics[width=\linewidth]{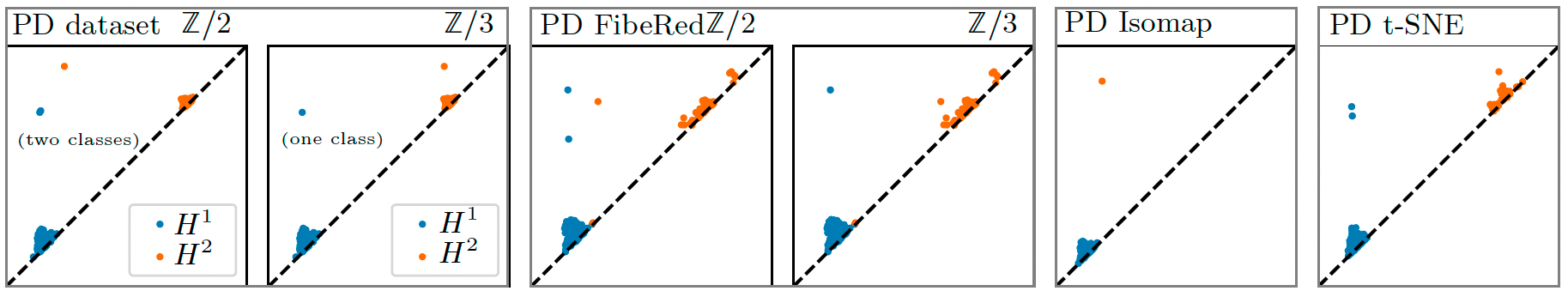}
    \end{center}
    \caption{The PD of the original data with $\Z/2$ (two prominent $1$-dimensional and one $2$-dimensional classes) and $\Z/3$ coefficients (one prominent $1$-dimensional class), which suggests the data is a Klein bottle;
    the PD of the representation obtained using \fibred{}, which matches the original topology well;
    the PD of a representation using Isomap;
    the PD of a representation using t-SNE.
    For Isomap and t-SNE, the PD is the same regardless of the field of coefficients. }
    \label{figure:klein-example}
\end{figure}

\begin{figure}
    \begin{center}
        \includegraphics[width=\linewidth]{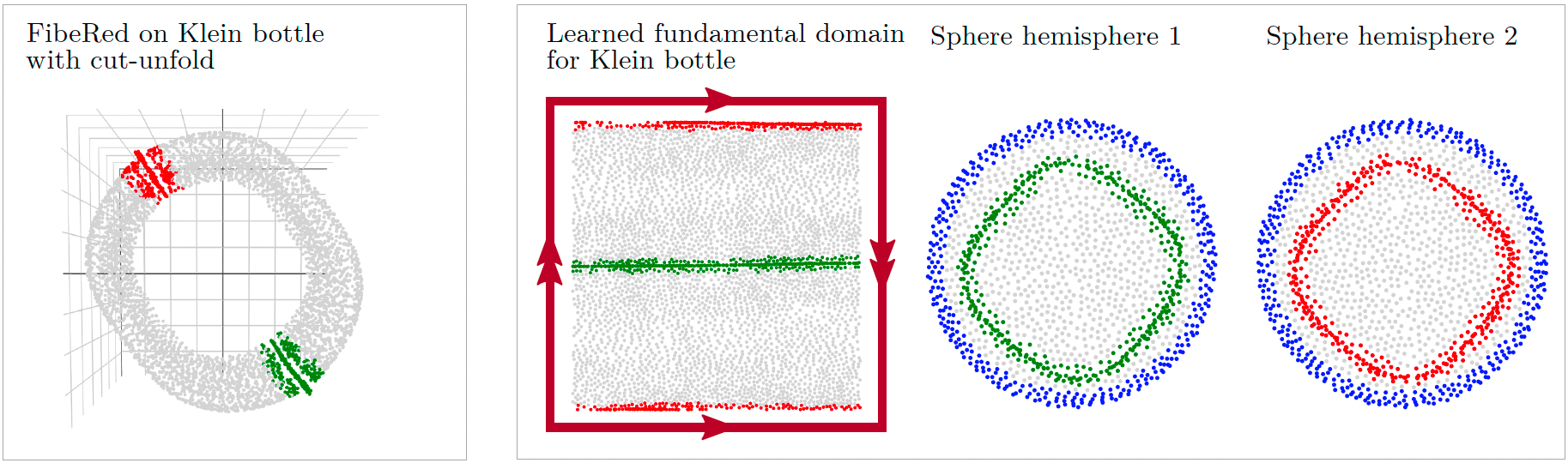}
    \end{center}
    \caption{\textit{Left}: The output of \fibred{} with the cut-unfold technique on the portion of the conformation space belonging to the Klein bottle.
    Colored in red and green are the two circles that glue the Klein bottle to the 2-sphere.
    Using this representation---a cylinder---we compute a new circular coordinate, which we combine with the initial circular coordinate to get a fundamental domain for the Klein bottle.
    \textit{Right}: A 2D model of the conformation space of cyclooctane.
    The two circles are two hemispheres of the 2D sphere and were obtained using Isomap.
    Points not colored in grey indicate the gluings that have to be performed to recover the conformation space.}
    \label{figure:cyclooctane-full-parametrization}
\end{figure}

By estimating the local dimension of the data, we first separate the Klein bottle part of the dataset from the 2-sphere.
In \cref{figure:klein-example}, we embed the Klein bottle part of the data in 4D.
We were not able to recover the right topology in $\R^4$ using any of LE, DM, LLE, HLLE, LTSA, Isomap, or t-SNE.
Meanwhile, UMAP is able to recover the right topology in $\R^4$.
In order to evaluate the 4D embeddings, we use the following distinguishing feature of the Klein bottle $K$: with $\Z/2$ coefficients we have $\dim(H^1(K;\Z/2)) = 2$ and $\dim(H^2(K;\Z/2))=1$, while with $\Z/3$ coefficients we have $\dim(H^1(K;\Z/3)) = 1$ and $\dim(H^2(K;\Z/3))=0$.
In \cref{figure:cyclooctane-full-parametrization}, we produce an efficient 2D parametrization of the conformation space of cyclooctane without using a priori knowledge of how the data was generated.

The difficulty faced by metric-based algorithms in this example is that the Klein bottle in high dimensional space has aspect ratio close to $1$ (i.e., an isometric representation by a fundamental domain such as the one \cref{figure:cyclooctane-full-parametrization} (left) has commensurable height and width), and thus it does not admit a simple isometric embedding in $\R^4$.

\section{Discussion}
\label{section:discussion}

We have presented a procedure to learn vector bundles from data and demonstrated that it can be used to decouple the global topology from the local geometry in topologically complex data.
We showed with examples that this can be helpful for embedding topologically complex data in low dimension, as well as for charting such data.
We have also developed a mathematical foundation for this point of view.

\medskip\noindent\textbf{Limitations.}
The theory and methods presented in this paper assume that the data lives in the total space of a vector bundle.
There are two main ways in which real data can deviate from these assumptions: (1) There are singularities in the data manifold and thus the base map is not a vector bundle since fibers may have different dimensions; (2) the data contains outliers and only a core subset of the data satisfies the assumptions.
Two other important caveats are that (3) the procedure assumes that a base map is given and that (4) success depends on the first step of the procedure finding a good cover of the data.
We comment on these remarks below.

\medskip\noindent\textbf{Future work.}
With respect to (1), the situation in which the fibers of the base map can have different dimensions can be abstracted using the theory of stratified vector bundles \cite{baues-ferrario,baues-ferrario2}.
We believe that the main algorithm of \cref{section:main-algorithm} can be adjusted to account for different local dimensions by allowing the cocycle $\Omega$ between patches with different dimension to be a matrix in a Stiefel manifold instead of an orthogonal matrix.
With respect to (2), our procedures are robust with respect to limited amount of noise and the problem of devising extensions robust to outliers is left as future work.

With respect to (3), there are several ways to obtain non-linear initial representations.
First, other cohomological coordinates besides circular coordinates have been developed \cite{perea-projective,polanco-perea}.
Second, one could use standard non-linear representations, such as the ones learned by Diffusion Maps \cite{lafon-lee,coifman-lafon}.
Third, one could use any of the lens functions \cite[Section~4]{singh-memoli-carlsson} Mapper uses.
Another interesting avenue for constructing coarse topological representations is to build a graph on the data, simplify it while preserving part of its large scale topology, and use a graph layout algorithm.
Regarding (4), finding good covers of noisy data is an interesting problem in itself; we believe the approach presented in this paper can be made more robust by developing a more nuanced subroutine for computing a cover.

Our approach depends on several constructions, some of which are known to be consistent estimators.
Addressing the consistency of the entire pipeline is left for future work.

Finally, \fibred{} can be interpreted as principal component analysis relative to an initial representation, as it works by linearly embedding the local coordinates of $X$ that are not already accounted by the initial map, in a way that is globally consistent and orthogonal to the coordinates already accounted by the initial map.
This suggests considering versions of other popular dimensionality reduction algorithms relative to an initial representation.

\bibliography{bibliography}

\begin{thebibliography}{10}

\bibitem{aamari-kim-chazal-michel-rinaldo-wasserman}
Eddie Aamari, Jisu Kim, Fr{\'e}d{\'e}ric Chazal, Bertrand Michel, Alessandro
  Rinaldo, and Larry Wasserman.
\newblock Estimating the reach of a manifold.
\newblock {\em Electronic journal of statistics}, 13(1):1359--1399, 2019.

\bibitem{abarbanel}
Henry D.~I. Abarbanel.
\newblock {\em Analysis of observed chaotic data}.
\newblock Institute for Nonlinear Science. Springer-Verlag, New York, 1996.
\newblock \href {https://doi.org/10.1007/978-1-4612-0763-4}
  {\path{doi:10.1007/978-1-4612-0763-4}}.

\bibitem{balabin}
Roman~M Balabin.
\newblock Enthalpy difference between conformations of normal alkanes: Raman
  spectroscopy study of n-pentane and n-butane.
\newblock {\em The Journal of Physical Chemistry A}, 113(6):1012--1019, 2009.

\bibitem{bandeira-singer-spielman}
Afonso~S. Bandeira, Amit Singer, and Daniel~A. Spielman.
\newblock A {C}heeger inequality for the graph connection {L}aplacian.
\newblock {\em SIAM J. Matrix Anal. Appl.}, 34(4):1611--1630, 2013.
\newblock \href {https://doi.org/10.1137/120875338}
  {\path{doi:10.1137/120875338}}.

\bibitem{bauer}
Ulrich Bauer.
\newblock Ripser: efficient computation of vietoris-rips persistence barcodes.
\newblock {\em Journal of Applied and Computational Topology}, 2021.
\newblock \href {https://doi.org/10.1007/s41468-021-00071-5}
  {\path{doi:10.1007/s41468-021-00071-5}}.

\bibitem{baues-ferrario2}
Hans-Joachim Baues and Davide~L. Ferrario.
\newblock {$K$}-theory of stratified vector bundles.
\newblock {\em $K$-Theory}, 28(3):259--284, 2003.
\newblock URL: \url{https://doi-org.ezproxy.neu.edu/10.1023/A:1026215632002},
  \href {https://doi.org/10.1023/A:1026215632002}
  {\path{doi:10.1023/A:1026215632002}}.

\bibitem{baues-ferrario}
Hans-Joachim Baues and Davide~L. Ferrario.
\newblock Stratified fibre bundles.
\newblock {\em Forum Math.}, 16(6):865--902, 2004.
\newblock URL:
  \url{https://doi-org.ezproxy.neu.edu/10.1515/form.2004.16.6.865}, \href
  {https://doi.org/10.1515/form.2004.16.6.865}
  {\path{doi:10.1515/form.2004.16.6.865}}.

\bibitem{belkin-niyogi}
Mikhail Belkin and Partha Niyogi.
\newblock Laplacian eigenmaps for dimensionality reduction and data
  representation.
\newblock {\em Neural computation}, 15(6):1373--1396, 2003.

\bibitem{borg-groenen}
Ingwer Borg and Patrick J.~F. Groenen.
\newblock {\em Modern multidimensional scaling}.
\newblock Springer Series in Statistics. Springer, New York, second edition,
  2005.
\newblock Theory and applications.

\bibitem{brand}
Matthew Brand.
\newblock Charting a manifold.
\newblock In S.~Becker, S.~Thrun, and K.~Obermayer, editors, {\em Advances in
  Neural Information Processing Systems}, volume~15. MIT Press, 2002.
\newblock URL:
  \url{https://proceedings.neurips.cc/paper/2002/file/8929c70f8d710e412d38da624b21c3c8-Paper.pdf}.

\bibitem{cartan}
Elie~Joseph Cartan.
\newblock Sur la possibilit\'{e} de plonger un espace riemannien donn\'{e} dans
  un espace euclidien.
\newblock {\em Annales de la Soci\'{e}t\'{e} Polonaise de Mathématique}, 1928.

\bibitem{casewit-colwell-rappe}
CJ~Casewit, KS~Colwell, and AK~Rappe.
\newblock Application of a universal force field to organic molecules.
\newblock {\em Journal of the American chemical society}, 114(25):10035--10046,
  1992.

\bibitem{charo-artana-sciamarella}
Gisela~D. Char\'{o}, Guillermo Artana, and Denisse Sciamarella.
\newblock Topology of dynamical reconstructions from {L}agrangian data.
\newblock {\em Phys. D}, 405:132371, 12, 2020.
\newblock \href {https://doi.org/10.1016/j.physd.2020.132371}
  {\path{doi:10.1016/j.physd.2020.132371}}.

\bibitem{vankadara-luxburg}
Leena Chennuru~Vankadara and Ulrike von Luxburg.
\newblock Measures of distortion for machine learning.
\newblock {\em Advances in Neural Information Processing Systems}, 31, 2018.

\bibitem{coifman-lafon}
Ronald~R. Coifman and St\'{e}phane Lafon.
\newblock Diffusion maps.
\newblock {\em Appl. Comput. Harmon. Anal.}, 21(1):5--30, 2006.
\newblock \href {https://doi.org/10.1016/j.acha.2006.04.006}
  {\path{doi:10.1016/j.acha.2006.04.006}}.

\bibitem{silva-morozov-vejdemo}
Vin de~Silva, Dmitriy Morozov, and Mikael Vejdemo-Johansson.
\newblock Persistent cohomology and circular coordinates.
\newblock {\em Discrete Comput. Geom.}, 45(4):737--759, 2011.
\newblock \href {https://doi.org/10.1007/s00454-011-9344-x}
  {\path{doi:10.1007/s00454-011-9344-x}}.

\bibitem{donoho-grimes}
David~L. Donoho and Carrie Grimes.
\newblock Hessian eigenmaps: Locally linear embedding techniques for
  high-dimensional data.
\newblock {\em Proceedings of the National Academy of Sciences},
  100(10):5591--5596, 2003.
\newblock URL: \url{https://www.pnas.org/doi/abs/10.1073/pnas.1031596100},
  \href
  {http://arxiv.org/abs/https://www.pnas.org/doi/pdf/10.1073/pnas.1031596100}
  {\path{arXiv:https://www.pnas.org/doi/pdf/10.1073/pnas.1031596100}}, \href
  {https://doi.org/10.1073/pnas.1031596100}
  {\path{doi:10.1073/pnas.1031596100}}.

\bibitem{fernandez}
Ximena Fern\'{a}ndez.
\newblock Topology of fluid flows.
\newblock \url{https://github.com/ximenafernandez/topology_fluids}, 2022.

\bibitem{garcia-menchaca-menchaca-pomares-perez-lakouari}
Jesus Garcia-Diaz, Rolando Menchaca-Mendez, Ricardo Menchaca-Mendez, Saúl
  Pomares~Hernández, Julio~César Pérez-Sansalvador, and Noureddine Lakouari.
\newblock Approximation algorithms for the vertex k-center problem: Survey and
  experimental evaluation.
\newblock {\em IEEE Access}, 7:109228--109245, 2019.
\newblock \href {https://doi.org/10.1109/ACCESS.2019.2933875}
  {\path{doi:10.1109/ACCESS.2019.2933875}}.

\bibitem{ghrist}
Robert Ghrist.
\newblock Barcodes: the persistent topology of data.
\newblock {\em Bull. Amer. Math. Soc. (N.S.)}, 45(1):61--75, 2008.
\newblock \href {https://doi.org/10.1090/S0273-0979-07-01191-3}
  {\path{doi:10.1090/S0273-0979-07-01191-3}}.

\bibitem{haile}
James~M Haile.
\newblock {\em Molecular dynamics simulation: elementary methods}.
\newblock John Wiley \& Sons, Inc., 1992.

\bibitem{halpern-weaver}
B.~Halpern and C.~Weaver.
\newblock Inverting a cylinder through isometric immersions and isometric
  embeddings.
\newblock {\em Trans. Amer. Math. Soc.}, 230:41--70, 1977.
\newblock \href {https://doi.org/10.2307/1997711} {\path{doi:10.2307/1997711}}.

\bibitem{hatcher}
Allen Hatcher.
\newblock {\em Algebraic topology}.
\newblock Cambridge University Press, Cambridge, 2002.

\bibitem{horn-johnson}
Roger~A. Horn and Charles~R. Johnson.
\newblock {\em Matrix analysis}.
\newblock Cambridge University Press, Cambridge, second edition, 2013.

\bibitem{jost}
J\"{u}rgen Jost.
\newblock {\em Riemannian geometry and geometric analysis}.
\newblock Universitext. Springer, Cham, seventh edition, 2017.
\newblock \href {https://doi.org/10.1007/978-3-319-61860-9}
  {\path{doi:10.1007/978-3-319-61860-9}}.

\bibitem{lafon-lee}
S.~Lafon and A.B. Lee.
\newblock Diffusion maps and coarse-graining: a unified framework for
  dimensionality reduction, graph partitioning, and data set parameterization.
\newblock {\em IEEE Transactions on Pattern Analysis and Machine Intelligence},
  28(9):1393--1403, 2006.
\newblock \href {https://doi.org/10.1109/TPAMI.2006.184}
  {\path{doi:10.1109/TPAMI.2006.184}}.

\bibitem{lee-verleysen}
John~Aldo Lee and Michel Verleysen.
\newblock Nonlinear dimensionality reduction of data manifolds with essential
  loops.
\newblock {\em Neurocomput.}, 67:29--53, aug 2005.
\newblock \href {https://doi.org/10.1016/j.neucom.2004.11.042}
  {\path{doi:10.1016/j.neucom.2004.11.042}}.

\bibitem{little-lee-jung-maggioni}
Anna~V Little, Jason Lee, Yoon-Mo Jung, and Mauro Maggioni.
\newblock Estimation of intrinsic dimensionality of samples from noisy
  low-dimensional manifolds in high dimensions with multiscale svd.
\newblock In {\em 2009 IEEE/SP 15th Workshop on Statistical Signal Processing},
  pages 85--88. IEEE, 2009.

\bibitem{luo-xu-zhang-jin}
Zixiang Luo, Chenyu Xu, Zhen Zhang, and Wenfei Jin.
\newblock A topology-preserving dimensionality reduction method for single-cell
  rna-seq data using graph autoencoder.
\newblock {\em Scientific reports}, 11(1):1--8, 2021.

\bibitem{martin-thompson-coutsias-watson}
Shawn Martin, Aidan Thompson, Evangelos~A Coutsias, and Jean-Paul Watson.
\newblock Topology of cyclo-octane energy landscape.
\newblock {\em The journal of chemical physics}, 132(23):234115, 2010.

\bibitem{mcinnes-healy-melville}
Leland McInnes, John Healy, and James Melville.
\newblock Umap: Uniform manifold approximation and projection for dimension
  reduction, 2018.
\newblock URL: \url{https://arxiv.org/abs/1802.03426}, \href
  {https://doi.org/10.48550/ARXIV.1802.03426}
  {\path{doi:10.48550/ARXIV.1802.03426}}.

\bibitem{membrillo-pirashvili-steinberg-brodzki-frey}
Ingrid Membrillo-Solis, Mariam Pirashvili, Lee Steinberg, Jacek Brodzki, and
  Jeremy~G. Frey.
\newblock Topology and geometry of molecular conformational spaces and energy
  landscapes, 2019.
\newblock URL: \url{https://arxiv.org/abs/1907.07770}, \href
  {https://doi.org/10.48550/ARXIV.1907.07770}
  {\path{doi:10.48550/ARXIV.1907.07770}}.

\bibitem{milnor-stasheff}
John~W. Milnor and James~D. Stasheff.
\newblock {\em Characteristic classes}.
\newblock Annals of Mathematics Studies, No. 76. Princeton University Press,
  Princeton, N. J.; University of Tokyo Press, Tokyo, 1974.

\bibitem{moor-horn-rick-borgwardt}
Michael Moor, Max Horn, Bastian Rieck, and Karsten Borgwardt.
\newblock Topological autoencoders.
\newblock In {\em International conference on machine learning}, pages
  7045--7054. PMLR, 2020.

\bibitem{oudot}
Steve~Y. Oudot.
\newblock {\em Persistence theory: from quiver representations to data
  analysis}, volume 209 of {\em Mathematical Surveys and Monographs}.
\newblock American Mathematical Society, Providence, RI, 2015.
\newblock \href {https://doi.org/10.1090/surv/209}
  {\path{doi:10.1090/surv/209}}.

\bibitem{paul-chalup}
Rahul Paul and Stephan~K Chalup.
\newblock A study on validating non-linear dimensionality reduction using
  persistent homology.
\newblock {\em Pattern Recognition Letters}, 100:160--166, 2017.

\bibitem{scikit-learn}
F.~Pedregosa, G.~Varoquaux, A.~Gramfort, V.~Michel, B.~Thirion, O.~Grisel,
  M.~Blondel, P.~Prettenhofer, R.~Weiss, V.~Dubourg, J.~Vanderplas, A.~Passos,
  D.~Cournapeau, M.~Brucher, M.~Perrot, and E.~Duchesnay.
\newblock Scikit-learn: Machine learning in {P}ython.
\newblock {\em Journal of Machine Learning Research}, 12:2825--2830, 2011.

\bibitem{perea-projective}
Jose~A. Perea.
\newblock Multiscale projective coordinates via persistent cohomology of sparse
  filtrations.
\newblock {\em Discrete Comput. Geom.}, 59(1):175--225, 2018.
\newblock \href {https://doi.org/10.1007/s00454-017-9927-2}
  {\path{doi:10.1007/s00454-017-9927-2}}.

\bibitem{perea}
Jose~A. Perea.
\newblock Topological times series analysis.
\newblock {\em Notices Amer. Math. Soc.}, 66(5):686--694, 2019.

\bibitem{perea-circular}
Jose~A. Perea.
\newblock Sparse circular coordinates via principal {$\mathbb Z$}-bundles.
\newblock In {\em Topological data analysis---the {A}bel {S}ymposium 2018},
  volume~15 of {\em Abel Symp.}, pages 435--458. Springer, Cham, 2020.

\bibitem{polanco-perea}
Luis Polanco and Jose~A. Perea.
\newblock Coordinatizing data with lens spaces and persistent cohomology.
\newblock In Zachary Friggstad and Jean{-}Lou~De Carufel, editors, {\em
  Proceedings of the 31st Canadian Conference on Computational Geometry, {CCCG}
  2019, August 8-10, 2019, University of Alberta, Edmonton, Alberta, Canada},
  pages 49--58, 2019.

\bibitem{rdkit}
{RDK}it developers.
\newblock {RDK}it: {O}pen-source cheminformatics.
\newblock \url{http://www.rdkit.org}, 2022.

\bibitem{rieck-leitte}
Bastian Rieck and Heike Leitte.
\newblock Persistent homology for the evaluation of dimensionality reduction
  schemes.
\newblock In {\em Computer Graphics Forum}, volume~34, pages 431--440. Wiley
  Online Library, 2015.

\bibitem{roweis-saul-hinton}
Sam Roweis, Lawrence Saul, and Geoffrey~E Hinton.
\newblock Global coordination of local linear models.
\newblock In T.~Dietterich, S.~Becker, and Z.~Ghahramani, editors, {\em
  Advances in Neural Information Processing Systems}, volume~14. MIT Press,
  2001.
\newblock URL:
  \url{https://proceedings.neurips.cc/paper/2001/file/850af92f8d9903e7a4e0559a98ecc857-Paper.pdf}.

\bibitem{roweis-saul}
Sam~T Roweis and Lawrence~K Saul.
\newblock Nonlinear dimensionality reduction by locally linear embedding.
\newblock {\em science}, 290(5500):2323--2326, 2000.

\bibitem{sadeghi-ghasemi-schaefer-mohr-lill-goedecker}
Ali Sadeghi, S~Alireza Ghasemi, Bastian Schaefer, Stephan Mohr, Markus~A Lill,
  and Stefan Goedecker.
\newblock Metrics for measuring distances in configuration spaces.
\newblock {\em The Journal of chemical physics}, 139(18):184118, 2013.

\bibitem{scoccola2023}
Luis Scoccola, Hitesh Gakhar, Johnathan Bush, Nikolas Schonsheck, Tatum Rask,
  Ling Zhou, and Jose~A. Perea.
\newblock Toroidal coordinates: Decorrelating circular coordinates with lattice
  reduction, 2022.
\newblock URL: \url{https://arxiv.org/abs/2212.07201}, \href
  {https://doi.org/10.48550/ARXIV.2212.07201}
  {\path{doi:10.48550/ARXIV.2212.07201}}.

\bibitem{implementation}
Luis Scoccola and Jose~A. Perea.
\newblock Fibe{R}ed implementation.
\newblock \url{https://github.com/LuisScoccola/fibered.git}, 2022.

\bibitem{scoccola-perea}
Luis Scoccola and Jose~A. Perea.
\newblock Approximate and discrete {E}uclidean vector bundles.
\newblock {\em Forum of Mathematics, Sigma (to appear)}, 2023.

\bibitem{shadden-lekien-marsden}
Shawn~C. Shadden, Francois Lekien, and Jerrold~E. Marsden.
\newblock Definition and properties of {L}agrangian coherent structures from
  finite-time {L}yapunov exponents in two-dimensional aperiodic flows.
\newblock {\em Phys. D}, 212(3-4):271--304, 2005.
\newblock \href {https://doi.org/10.1016/j.physd.2005.10.007}
  {\path{doi:10.1016/j.physd.2005.10.007}}.

\bibitem{singer}
A.~Singer.
\newblock Angular synchronization by eigenvectors and semidefinite programming.
\newblock {\em Appl. Comput. Harmon. Anal.}, 30(1):20--36, 2011.
\newblock \href {https://doi.org/10.1016/j.acha.2010.02.001}
  {\path{doi:10.1016/j.acha.2010.02.001}}.

\bibitem{singer-wu}
Amit Singer and Hau-tieng Wu.
\newblock Orientability and diffusion maps.
\newblock {\em Appl. Comput. Harmon. Anal.}, 31(1):44--58, 2011.
\newblock \href {https://doi.org/10.1016/j.acha.2010.10.001}
  {\path{doi:10.1016/j.acha.2010.10.001}}.

\bibitem{singh-memoli-carlsson}
Gurjeet Singh, Facundo M{\'e}moli, Gunnar~E Carlsson, et~al.
\newblock Topological methods for the analysis of high dimensional data sets
  and 3d object recognition.
\newblock {\em PBG Eurographics}, 2, 2007.

\bibitem{tang-liu-zhang-mei}
Jian Tang, Jingzhou Liu, Ming Zhang, and Qiaozhu Mei.
\newblock Visualizing large-scale and high-dimensional data.
\newblock In {\em Proceedings of the 25th International Conference on World
  Wide Web}, WWW '16, pages 287--297, Republic and Canton of Geneva, CHE, 2016.
  International World Wide Web Conferences Steering Committee.
\newblock \href {https://doi.org/10.1145/2872427.2883041}
  {\path{doi:10.1145/2872427.2883041}}.

\bibitem{teh-roweis}
Yee Teh and Sam Roweis.
\newblock Automatic alignment of local representations.
\newblock In S.~Becker, S.~Thrun, and K.~Obermayer, editors, {\em Advances in
  Neural Information Processing Systems}, volume~15. MIT Press, 2002.
\newblock URL:
  \url{https://proceedings.neurips.cc/paper/2002/file/3a1dd98341fafc1dfe9bcf36360e6b84-Paper.pdf}.

\bibitem{tenenbaum-silva-langford}
Joshua~B Tenenbaum, Vin~de Silva, and John~C Langford.
\newblock A global geometric framework for nonlinear dimensionality reduction.
\newblock {\em Science}, 290(5500):2319--2323, 2000.

\bibitem{dreimac}
Chris Tralie, Tom Mease, and Jose Perea.
\newblock Dreimac.
\newblock \url{https://github.com/ctralie/DREiMac}, 2017.

\bibitem{maaten-hinton}
Laurens Van~der Maaten and Geoffrey Hinton.
\newblock Visualizing data using t-{SNE}.
\newblock {\em Journal of machine learning research}, 9(11), 2008.

\bibitem{wagner-solomon-bendich}
Alexander Wagner, Elchanan Solomon, and Paul Bendich.
\newblock Improving metric dimensionality reduction with distributed topology,
  2021.
\newblock URL: \url{https://arxiv.org/abs/2106.07613}, \href
  {https://doi.org/10.48550/ARXIV.2106.07613}
  {\path{doi:10.48550/ARXIV.2106.07613}}.

\bibitem{whitney}
Hassler Whitney.
\newblock The self-intersections of a smooth {$n$}-manifold in {$2n$}-space.
\newblock {\em Ann. of Math. (2)}, 45:220--246, 1944.
\newblock URL: \url{https://doi-org.ezproxy.neu.edu/10.2307/1969265}, \href
  {https://doi.org/10.2307/1969265} {\path{doi:10.2307/1969265}}.

\bibitem{yan-zhao-rosen-scheidegger-wang}
Lin Yan, Yaodong Zhao, Paul Rosen, Carlos Scheidegger, and Bei Wang.
\newblock Homology-preserving dimensionality reduction via manifold landmarking
  and tearing.
\newblock {\em arXiv preprint arXiv:1806.08460}, 2018.

\bibitem{zhang-zha}
Zhenyue Zhang and Hongyuan Zha.
\newblock Principal manifolds and nonlinear dimensionality reduction via
  tangent space alignment.
\newblock {\em SIAM Journal on Scientific Computing}, pages 313--338, 2004.

\end{thebibliography}

\newpage

\appendix

\section{Theory}

\subsection{Background and notation}
\label{section:background}

We give a---necessarily terse and sometimes informal---description of the main topological notions relevant to this paper. 
We include detailed references for the interested reader.

\medskip
\noindent\textbf{Vector bundles.}
We assume that manifolds, vector bundles, maps, and metrics are all smooth, i.e., $C^\infty$.
For an introduction, we refer the reader to, e.g., \cite{milnor-stasheff,jost}.

A \define{cover} of a topological space $\B$ is an indexed collection $\UUU = \{U_i\}_{i \in I}$ of open sets of $\B$ such that $\B = \bigcup_{i \in I} U_i$.
When there is no risk of confusion, we may omit the indexing set.

A rank $r$ (smooth) \define{vector bundle} $\pi : \X \to \B$ consists of a smooth map of differentiable manifolds such that each fiber $\pi^{-1}(b) \subseteq \X$ is endowed with the structure of a dimension $r$ real vector space, and such that there exists a cover $\UUU = \{U_i\}$ of $\B$ and diffeomorphisms $(\pi|_{\pi^{-1}(U_i)},f_i) : \pi^{-1}(U_i) \to U_i \times \R^r$ that induce a linear isomorphism $f_i|_{\pi^{-1}(b)} : \pi^{-1}(b) \to \R^r$ for each $b \in U_i \subseteq B$.
The spaces $\X$ and $\B$ are called the \define{total space} and the \define{base space}, respectively,
and the maps $\{(\pi|_{\pi^{-1}(U_i)},f_i) : \pi^{-1}(U_i) \to U_i \times \R^r\}$ are called a \define{trivialization} of the vector bundle $\pi$.

A vector bundle $\pi : \X \to \B$ is \define{Euclidean} if each of its fibers $\pi^{-1}(b)$ is endowed with a scalar product $\langle -,-\rangle_b$ that varies smoothly with $b \in B$.
A \define{metric trivialization} of a Euclidean vector bundle $\pi : \X \to \B$ consists of a trivialization in which the induced linear isomorphisms $\pi^{-1}(b) \to \R^r$ are isometries, where $\R^r$ is endowed with its usual scalar product.

A \define{Riemannian manifold} consists of a smooth manifold $\B$ together with a Euclidean vector bundle structure on its tangent vector bundle $T\B \to \B$.

Let $n \leq m$.
The (compact) \define{Stiefel manifold} $\stief(n,m)$ consists of the space of $m$-by-$n$ matrices with orthonormal columns, endowed with its usual differentiable structure.
When $n=m$, the Stiefel manifold $\stief(n,n)$ is equal to the \define{orthogonal group} $O(n)$ of orthogonal $n$-by-$n$ matrices.

A \define{partition of unity} subordinate to a cover $\UUU = \{U_i\}$ of a topological space $\B$ consists of a family of continuous functions $\{\rho_i : \B \to \R\}$ taking non-negative values such that $\rho_i(x) = 0$ if $x \notin U_i$, for all $x \in \B$ we have that $\rho_i(x) \neq 0$ for only finitely many $i \in I$, and such that $\sum_{i} \rho_i(x) = 1$ for all $x \in \B$.

\medskip
\noindent\textbf{Simplicial complexes.}
A \define{simplicial complex} consists of a set $S$ together with a family $\simp(S)$ of non-empty, finite subsets $S$, called \define{simplices}, that is closed under taking subsets and that contains all singletons.
Any graph $G$ gives rise to a simplicial complex whose underlying set is the set of vertices of $G$, and whose simplices consist of the vertices and the edges of $G$.
Another important example is that of the \define{nerve} of a cover $\UUU = \{U_i\}_{i \in I}$ of a topological space.
The underlying set of the nerve of $\UUU$ is $I$, while the simplices consist of all finite subsets $J \subseteq I$ such that $\bigcap_{j \in J} U_j \neq \emptyset$.

\medskip
\noindent\textbf{(Persistent) cohomology.}
We briefly recall some of the properties of cohomology \cite{hatcher} and persistence \cite{ghrist}.
Given a topological space $\B$, a field $\field$, and $n \in \N$, the \define{$n$th cohomology group} of $\B$ with coefficients in $\field$ is a $\field$-vector space $H^n(\B;\field)$, whose dimension, informally, counts the number of $n$-dimensional wholes in $\B$.
Cohomology is a functorial operation, which in particular implies that, given a family of topological spaces $\{\B_s\}_{s \in \R}$ such that $\B_s \subseteq \B_{s'}$ for $s \leq s'$, there exist linear maps $(s\leq s')^* : H^n(\B_{s'};\field) \to H^n(\B_{s};\field)$ such that $(s\leq s')^* \circ (s' \leq s'')^* = (s \leq s'')^*$ for all $s \leq s' \leq s'' \in \R$.

Under mild hypothesis, the family of $\field$-vector spaces and linear maps $\{H^n(\B_{s'};\field) \to H^n(\B_{s};\field)\}$ can be described by a \define{persistence diagram} (PD) \cite[Chapter~1,~Section~3]{oudot}, which consists of a finite multiset of points $\{(x_i,y_i) \in \R^2 : y_i > x_i\}$, which has the property that the rank of the linear map $H^n(\B_{s'};\field) \to H^n(\B_{s};\field)$ is equal to the number of points $(x_i, y_i)$ such that $x_i \leq s \leq s' < y_i$.
Informally, a point $(x_i, y_i)$ in the $n$th persistence diagram of a filtered topological space $\{\B_s\}_{s \in \R}$ represents a hole in the filtration that first appears in $\B_{x_i}$ and disappears (is filled) in $\B_{y_i}$.

When $\{\B_s\}$ is a well-behaved filtration, such a filtration of a finite simplicial complex, persistent cohomology can be used to construct cohomological coordinates, which are maps from the space that is being filtered into a topologically interesting space \cite{silva-morozov-vejdemo,perea-circular,perea-projective,polanco-perea}.
We use \define{circular coordinates} as in \cite{perea-circular}, which, given a choice of point in the persistence diagram of the first cohomology group of the Vietoris--Rips filtration \cite[Definition~1.2]{ghrist} of (a subsample of) the data, returns a map from the data into the circle.
The main takeaway here is that circular coordinates give a map into the circle which captures circularity present in the data.

\medskip
\noindent\textbf{Some spaces with non-trivial topology.}
We describe the torus, the M\"obius band, and the Klein bottle.
Given a topological space $\B$ and an equivalence relation $\sim$ on (the underlying set of) $\B$, the \define{quotient topology} on the set $\B/\sim$ is the topology defined by letting a set of $\B/\sim$ be open if and only if its preimage along the quotient map $\B \to \B/\sim$ is open in $\B$.

For example, as a topological space, the circle $S^1$ is the quotient of the interval $[0,1]$ with its usual topology, by the equivalence relation generated by $0 \sim 1$.
The subset $[0,1)$ of the interval is a fundamental domain for the circle.

The \define{M\"obius band} is the quotient of the square $[0,1] \times [0,1]$ by the equivalence relation generated by letting $(0,y) \sim (1,1-y)$ for all $y \in [0,1]$.
The subset $[0,1) \times [0,1]$ of the square is a fundamental domain for the M\"obius band.

The \define{torus} is the quotient of the square $[0,1] \times [0,1]$ by the equivalence relation generated by letting $(x,0) \sim (x,1)$ for all $x \in [0,1]$ and $(0,y) \sim (1,y)$ for all $y \in [0,1]$.
The subset $[0,1) \times [0,1)$ is a fundamental domain for the torus.

The \define{Klein bottle} is the quotient of the square $[0,1] \times [0,1]$ by the equivalence relation generated by letting $(x,0) \sim (x,1)$ for all $x \in [0,1]$ and $(0,y) \sim (1,1-y)$ for all $y \in [0,1]$.
The subset $[0,1) \times [0,1)$ is a fundamental domain for the Klein bottle.

\subsection{Proofs}
\label{section:proofs}

Since we use this in both proofs, note that, by definition of the normal bundle $N \to \B$, \cref{problem:main-problem} admits a solution $j : N \to \R^D$ when $\X = N$.
In particular, for each $b \in \B$, we have a bijection $j|_{\nu^{-1}(b)} : \nu^{-1}(b) \to j(\nu^{-1}(b))$.

\begin{proof}[Proof of \cref{lemma:reduction-main-problem}]
    We start with the ``if'' direction.
    Given a morphism $\X \to N$ that is injective and an isometry in each fiber gives, by postcomposition with $j$, a fiberwise isometric embedding $\X \to \R^D$ satisfying the requirements in the statement of \cref{problem:main-problem}.

    For the other direction, assume given an isometric embedding $\overline{\iota} : \X \to \R^D$ satisfying the requirements in the statement of \cref{problem:main-problem}.
    Since $\overline{\iota}$ extends $\iota$ and is orthogonal to $\B$, its image is contained in the image of $j : N \to \R^D$.
    The claim now follows from the fact that $\overline{\iota}$ and $j$ is are linear isometries in each fiber.
\end{proof}

\begin{proof}[Proof of \cref{proposition:cocycle-characterization}]
    We start with the ``if'' direction.
    Given the maps $\Phi = \{\Phi_i : U_i \to \stief(d,D-e)\}$ as in the statement, one defines a map $\X \to N$ by
    \[
        X_i \ni x \mapsto \left(j|_{\nu^{-1}(b)}\right)^{-1}\big(\alpha_i(\pi(x)) \Phi_i(\pi(x)) f_i(x) + \iota(\pi(x))\big) \in N,
    \]
    which, using \cref{equation:idealized-objective-function}, is easily seen to be a well-defined linear isometry in each fiber.

    For the other direction, suppose given a morphism of vector bundles $g : \X \to N$ that is injective and a linear isometry in each fiber.
    Using the fact that $g$ is linear, one sees that there exists a unique map $\Phi_i : U_i \to \stief(d,D-e)$ satisfying
    \[
        \Phi_i(\pi(x))^T \alpha_i(\pi(x))^T \big(j(g(x)) - \iota(\pi(x)))\big) = f_i(x),
    \]
    for all $x \in U_i$.
    This also implies, in particular, that the family of maps $\{\Phi_i\}$ satisfies \cref{equation:objective-function}.
\end{proof}

\section{Algorithm}
\subsection{Assumptions about input of \fibred{}}
\label{section:assumptions}

The assumptions here are made so that we can justify the steps of the algorithm; formally addressing the consistency of the algorithm is left for future work.

We assume that there exists a closed manifold $\B$ of dimension $e$, a rank $r$ Euclidean vector bundle $\pi : \X \to \B$, a Riemannian metric on $\X$,
and a smooth embedding $\iota : \B \to \R^D$.
We assume that the Riemannian metric on $\X$ is compatible with the Euclidean structure of the vector bundle $\pi$, in the sense that, given any metric trivialization of $\pi$ by maps $\{(\pi|_{\X_i},f_i) : \X_i \to U_i \times \R^r\}$, and $x \in \X$, the composite
\[
    \R^r \to T_{\pi(x)}\B \oplus T_{f_i(x)} \R^r \xrightarrow{\left(d (\pi|_{\X_i},f_i)_x\right)^{-1}} T_x \X
\]
is a linear isometric monomorphism.
Here, the first map is the inclusion in the second component of the direct sum, where we are using the standard identification between tangent spaces of a Euclidean space and the Euclidean space itself.

The input metric space of \fibred{} is assumed to be a subsample $X \subseteq \X$, which implies, in particular, that its image $B := \pi(X)$ is a subsample $B \subseteq \B$.

Note that we do not ask for the embedding $\B \subseteq \R^D$ to preserve the Riemannian structure of $\B$ inherited from that of $\X$.
Note also that, when solving the vector bundle embedding problem, only the Euclidean structure of $\pi$ is required to be preserved, although we assumed that the total space $\X$ has a (compatible) Riemannian metric.
We require a metric on $\X$ since most datasets come with a (global) distance, so we need to make some assumption about how this distance is generated.

\subsection{Efficiency}
\label{section:efficiency}
We comment on the complexity of the main subroutines of \fibred{}.
Let $N = |X|$.
The greedy approach to the $k$-center problem implemented by \covandpart{} has time complexity in $\mathcal{O}(kN)$.
\localrep{} and \estimatetangandnorm{} require $\mathcal{O}(k)$ SVD computations, with matrices of size $|X_i|$ for each $i$, which is at most $N$.
There is room for improvement here, since only eigenvectors corresponding to the $d$ and $D$ largest eigenvalues are required, respectively, and $|X_i|$ can be made significantly smaller than $N$ by making larger $k$.
Another significant improvement would be to use landmark MDS instead of MDS for \localrep{}.
\estimatefibcoord{} requires $\mathcal{O}(k)$ least squares solutions to a linear system, which can be implemented using SVD with matrices of size $e$.
\estimatereach{} requires $\mathcal{O}(k^2)$ direct calculations.
\estimatecocycles{} requires $\mathcal{O}(k^2)$ SVD.
Finally, \alignfibs{} requires $\mathcal{O}(\texttt{n\_iter})$ SVD with matrices of size $D-e$.

\section{Examples}
\subsection{Cut-unfold technique}
\label{section:cut-unfold}
Some topologically non-trivial spaces can be described as a simple quotient of a simple topological space.
For example, the circle $S^1$ is the quotient of the interval $[0,1]$ that identifies $0$ and $1$.
Thus, given a continuous map $\X \to S^1$, one gets a function $\X \to [0,1)$, which is still continuous if we ``cut'' the topology of $\X$ at the preimage of $0$.

\subsection{Pentane data generation}
\label{section:pentane-data-generation}
We used \texttt{RDKit}'s function \texttt{EmbedMolecule} with different random initializations to get a sample of $20000$ conformers of pentane.
We set \texttt{useExpTorsionAnglePrefs} to \texttt{False} in order to obtain a good sample of the full conformation space, rather than just conformers with low energy.
We then computed pairwise distances between all conformers using \texttt{GetBestRMS}, which computes the root-mean-square deviation (see, e.g., \cite{sadeghi-ghasemi-schaefer-mohr-lill-goedecker}).
In order to speed up the computation of pairwise distances, we compare the conformers after removing the hydrogen atoms and keeping the carbon atoms.
Finally, we kept a subsample of $5000$ conformers that approximates the full sample well.
The energy of conformers was calculated using universal force field \cite{casewit-colwell-rappe}.

\subsection{Computation of Stiefel--Whitney obstructions}
\label{section:obstructions-examples}

We describe how we compute Stiefel--Whitney classes associated to the vector bundles approximated by the cocycle $\Omega$ computed by \estimatecocycles{} in each of the three examples.
For this we use the algorithms of \cite[Theorem~C]{scoccola-perea} and follow the technique in \cite[Section~7.1]{scoccola-perea}.

By construction, $\Omega$ is a discrete approximate cocycle on the simplicial complex $\NNN$.
Recall that an edge $(ij)$ of $\NNN$ is associated the weight $s_{ij} = |X_i \cap X_j|$.
This determines a filtration of the simplicial complex $\NNN$, which we denote by $\{\NNN_{r}\}_{r \in [0,1]}$ where vertices are born at $0$, edges at $1 - s_{ij}/|X|$, and higher simplices are born when all of their edges appear.
In this way, $\NNN_{0}$ consists of just vertices and $\NNN_1 = \NNN$.
As in \cite[Section~7.1]{scoccola-perea}, we define the \define{death} of $\Omega$ as the supremum over all $r \in [0, 1]$ such that $\Omega$ is a $2$-approximate cocycle on $\NNN_r$, and the \define{span} of $\Omega$ as the region of the persistence diagram of classes that are born before the death of $\Omega$.
This is the region of classes that can appear in the support of $w_1(\Omega)$.

In \cref{figure:sw-obstructions}, in this document, we show the result of performing these computations in the three examples of \cref{section:examples}.

\subsection{Other runs}
\label{section:other-runs}

\begin{figure}
    \begin{center}
    \includegraphics[width=0.9\linewidth]{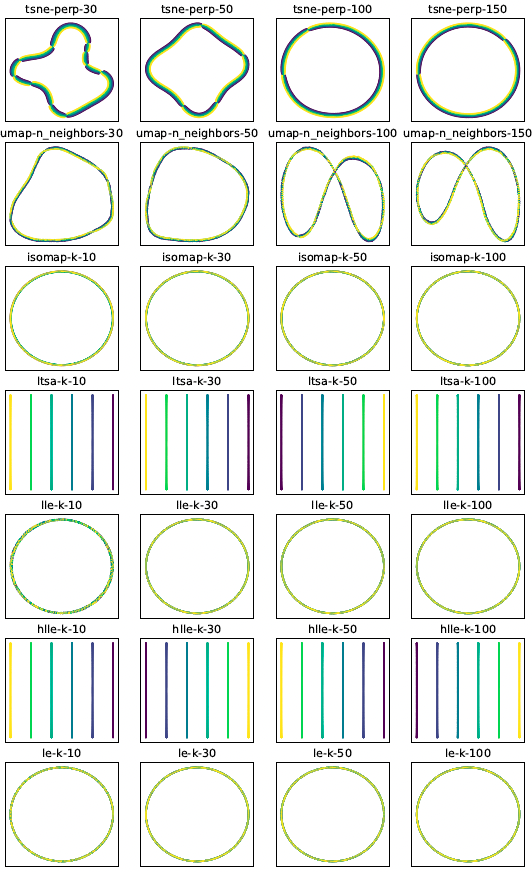}
    \end{center}
    \caption{
    We see that no output is able to consistently align the local 2D structure around the circularity of the cylinder.
    We do not include the results of running Diffusion Maps (the results are not significantly different from those of Laplacian Eigenmaps).}
    \label{figure:cylinder-embeddings}
\end{figure}

\begin{figure}
    \begin{center}
    \includegraphics[width=0.9\linewidth]{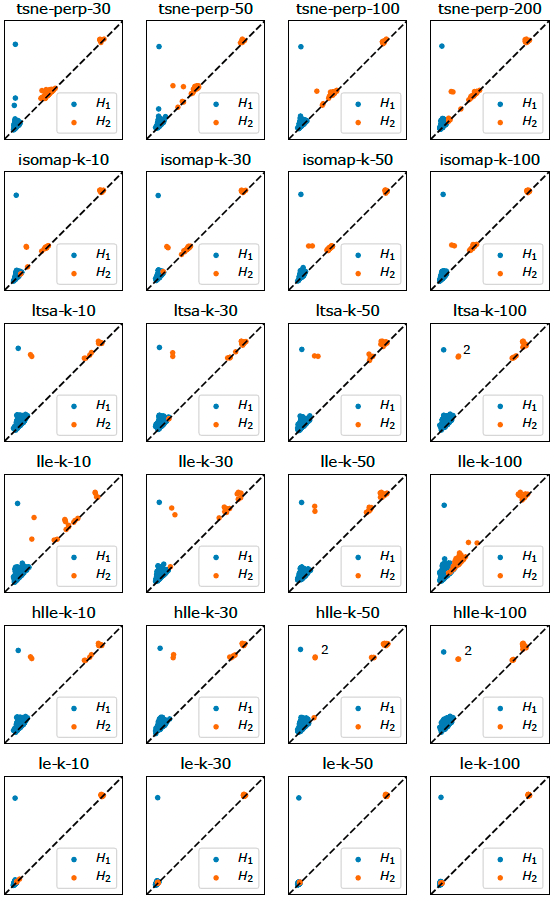}
    \end{center}
    \caption{Recall that the torus $T$ has homology $\dim(H_1(T;\Z/2)) = 2$ and $\dim(H_2(T;\Z/2)) = 1$.
    We see that no output has two prominent 1-dimensional classes and one prominent 2-dimensional class.
    Persistent homology was computed with $\Z/2$ coefficients.
    When there are two classes that almost overlap, we mark them with a $2$.
    We do not include the results of running Diffusion Maps (the results are not significantly different from those of Laplacian Eigenmaps) and UMAP (the results are not significantly different from those of t-SNE).}
    \label{figure:torus-pds}
\end{figure}

\begin{figure}
    \begin{center}
    \includegraphics[width=0.9\linewidth]{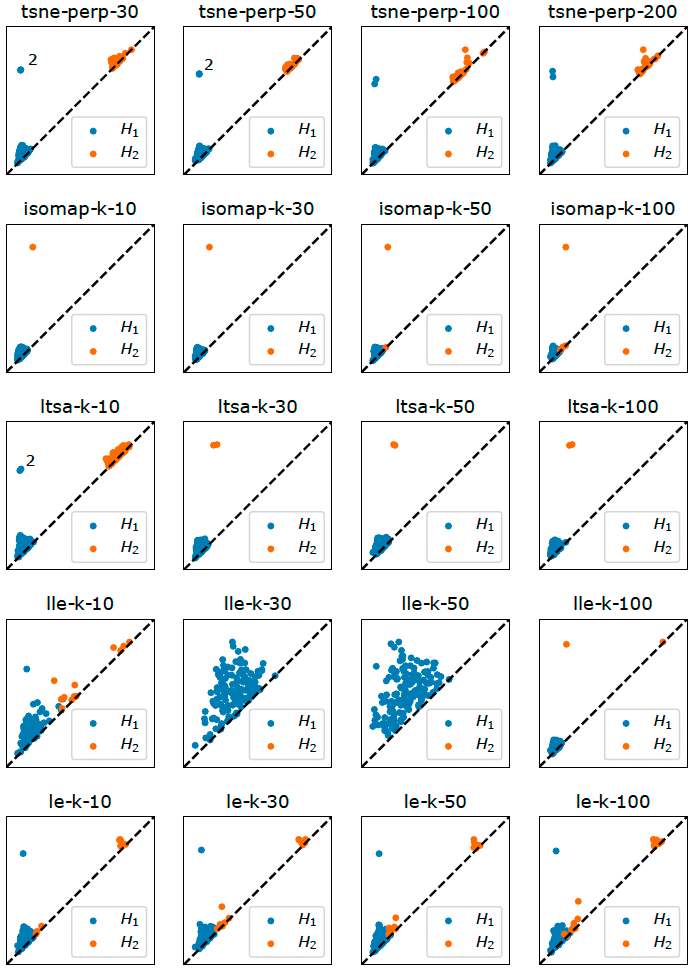}
    \end{center}
    \caption{Recall that, if $K$ is the Klein bottle, with $\Z/2$ coefficients we have $\dim(H_1(K;\Z/2)) = 2$ and $\dim(H_2(K;\Z/2))=1$, while with $\Z/3$ coefficients we have $\dim(H_1(K;\Z/3)) = 1$ and $\dim(H_2(K;\Z/3))=0$.
    We are displaying persistent homology with $\Z/2$ coefficients, which, for all the examples, coincides with the persistence homology with $\Z/3$ coefficients.
    We see that no output exhibits the homology of the Klein bottle.
    When there are two classes that almost overlap, we mark them with a $2$.
    In this example, HLLE runs fails when computing eigenvectors, since the data manifolds is not developable.
    We do not include the results of running Diffusion Maps (the results are not significantly different from those of Laplacian Eigenmaps).}
    \label{figure:klein-pds}
\end{figure}

We run Isomap, LLE, HLLE, LTSA, Laplacian Eigenmaps, Diffusion Maps, t-SNE, UMAP on the datasets of the paper.
We vary the main parameter for each algorithm.
We report here the results of the cylinder (\cref{figure:cylinder-embeddings}), torus (\cref{figure:torus-pds}), and Klein bottle data (\cref{figure:klein-pds}), as these can be judged by looking 2D plots.
We also report the minimal target dimension of each of the algorithms that recovers a topologically faithful embedding in \cref{table:minimal-target-dimensions}.
For t-SNE we use the PCA initialization.

As in the paper, we report persistent homology of a Vietoris--Rips complex on geodesic distance, computed as shortest path distance on a 15-nearest neighbor graph.

\begin{figure}
    \centering
    \begin{subfigure}[b]{0.25\textwidth}
        \centering
        \includegraphics[width=0.8\linewidth]{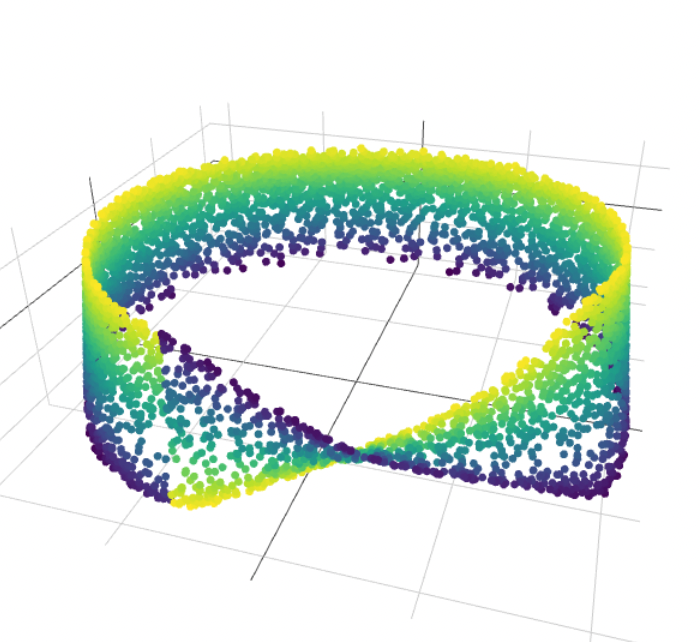}

        \includegraphics[width=0.8\linewidth]{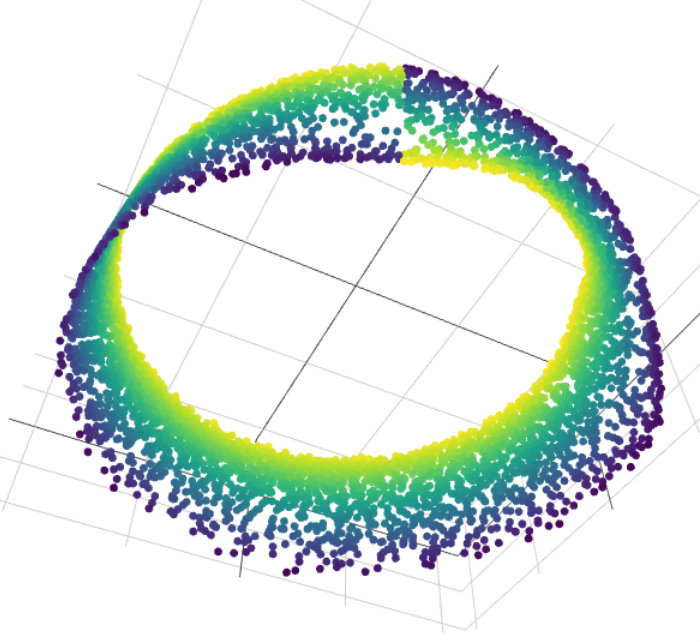}
        \caption{$k=10$}
        \label{fig:k-10}
    \end{subfigure}
    \hfill
    \begin{subfigure}[b]{0.25\textwidth}
        \centering
        \includegraphics[width=0.8\linewidth]{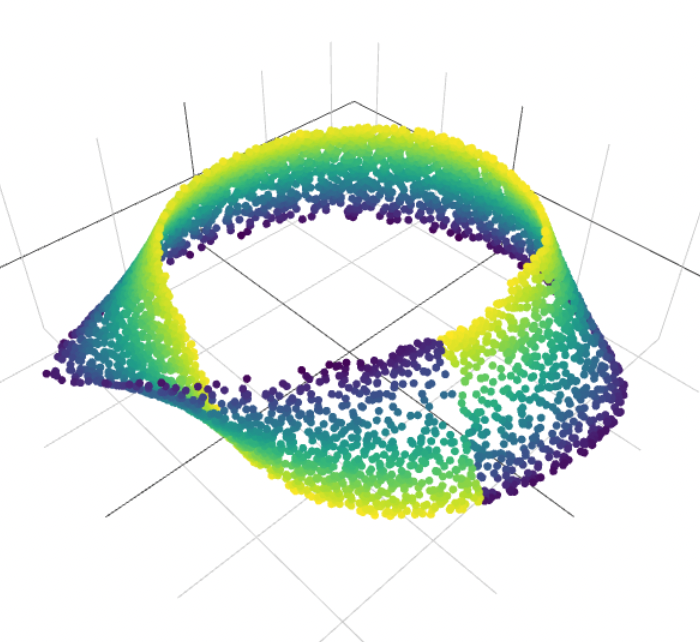}

        \includegraphics[width=0.8\linewidth]{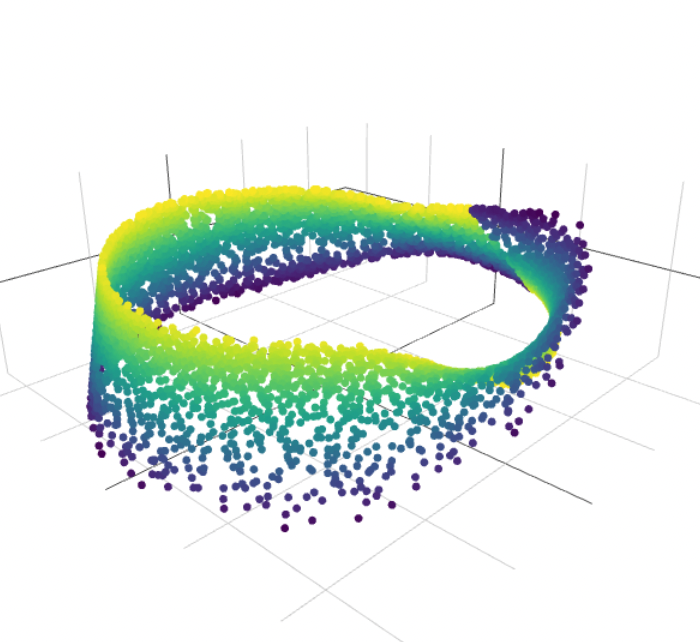}
        \caption{$k=32$}
        \label{fig:k-32}
    \end{subfigure}
    \hfill
    \begin{subfigure}[b]{0.25\textwidth}
        \centering
        \includegraphics[width=0.8\linewidth]{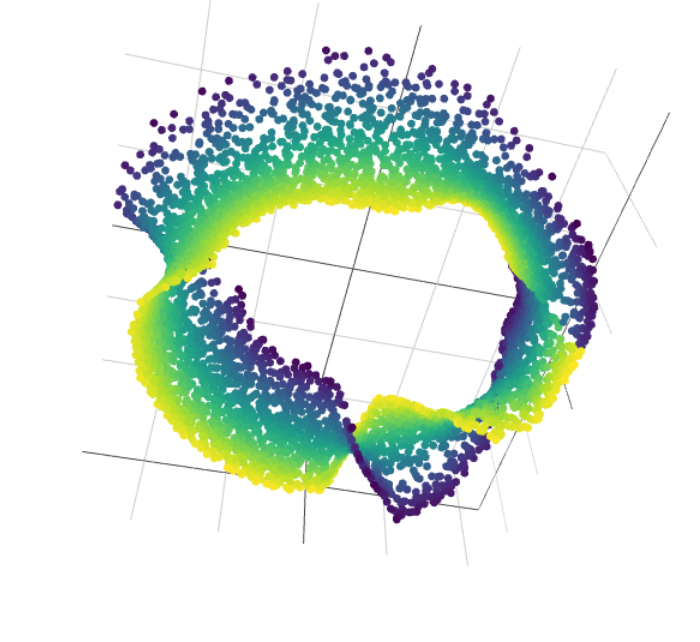}

        \includegraphics[width=0.8\linewidth]{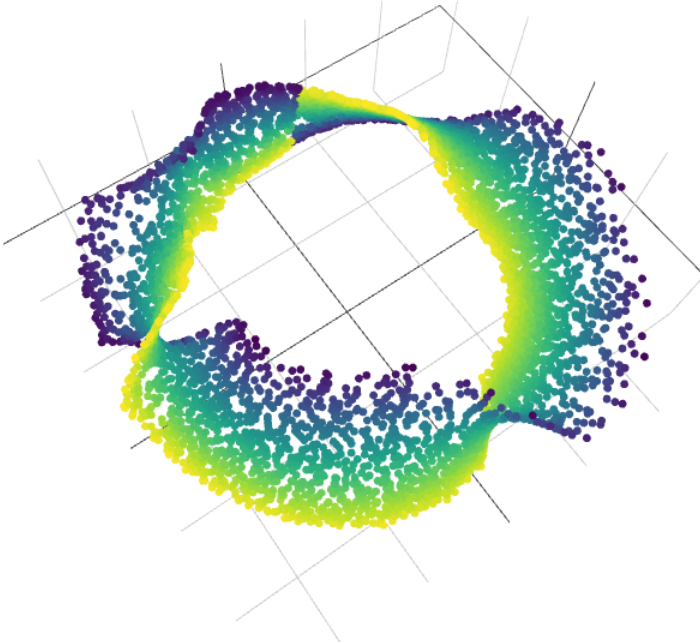}
 
        \caption{$k=48$}
        \label{fig:k48}
    \end{subfigure}
       \caption{\fibred{} on the M\"obius data for three choices of $k$.
       Choices between $k=10$ and $k=32$, containing, in particular, $k=16$, used in the main body of the paper, give smooth embeddings.
       We note that when the number of opens is small, $k=10$, topological changes (in this case the fiber changes orientation) tend to happen more abruptly.
       When the number of opens is large, $k \geq 32$, the objective function \cref{equation:objective-function} starts to have many local minima, since it only enforces local aligments.
       This is evident in the more extreme case $k=48$, where the technically topologically correct embedding is, nevertheless, quite irregular.
       This suggests that \cref{equation:objective-function} could be regularized in order to enforce more global alignments between the fibers.}
       \label{fig:cover-various-sizes}
\end{figure}

\begin{figure}
    \centering
    \begin{subfigure}[b]{0.23\textwidth}
        \centering
        \includegraphics[width=0.8\linewidth]{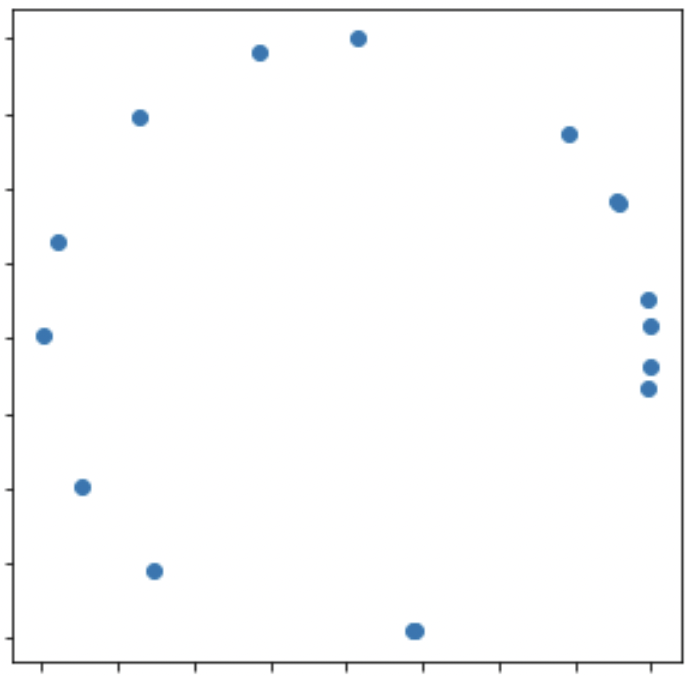}
        \caption{Centers for the cover.}
        \label{fig:cover-centers}
    \end{subfigure}
    \hspace{0.1cm}
     \begin{subfigure}[b]{0.23\textwidth}
        \includegraphics[width=\linewidth]{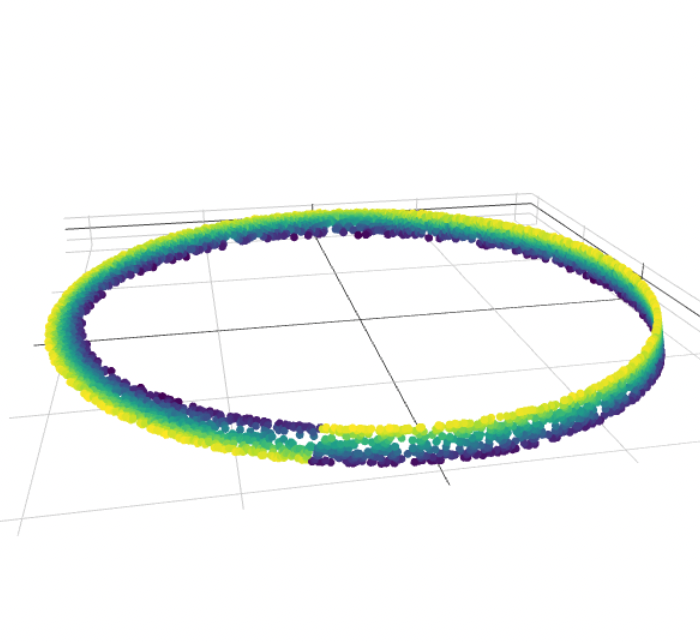}
        \caption{With random cover.}
        \label{fig:fibred-rand}
    \end{subfigure}
    \hspace{0.1cm}
     \begin{subfigure}[b]{0.46\textwidth}
        \includegraphics[width=0.49\linewidth]{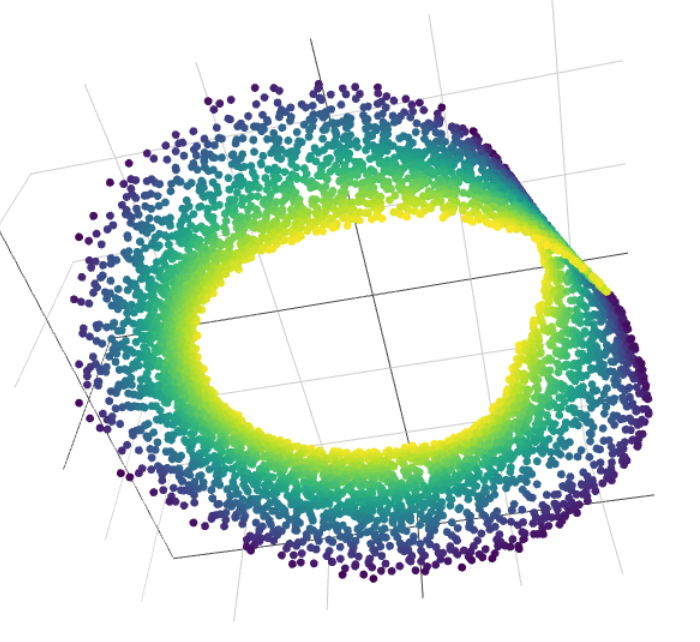}
        \includegraphics[width=0.49\linewidth]{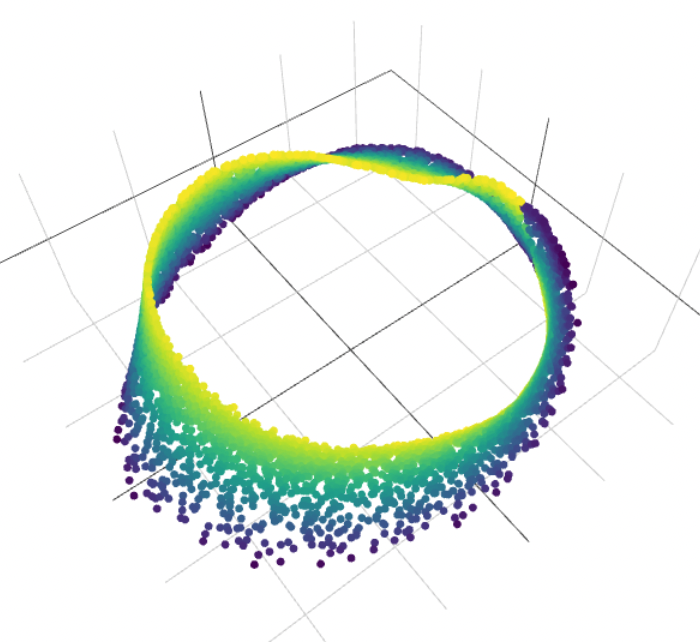}
        \caption{With random cover and adjusted reach.}
        \label{fig:fibred-rand-adjusted}
    \end{subfigure}
      \caption{We choose $16$ centers for the sets in the cover uniformly at random in the base space.
      Although the centers are not well distributed, the alignment of \fibred{} is successful.
      In \cref{fig:fibred-rand} we see that, without any adjustments, \fibred{} gives an embedding with very small fibers.
      This is explained by the fact that the reach computation gives a very small estimate, because of the large portion of the base space without any cover points.
      Nevertheless, by adjusting the reach manually, we see in \cref{fig:fibred-rand-adjusted} that the embedding is indeed topologically correct and quite regular.}
       \label{fig:cover-random}
\end{figure}

\begin{figure}
    \centering

    \begin{subfigure}[b]{1\textwidth}
        \centering
        \includegraphics[width=0.5\linewidth]{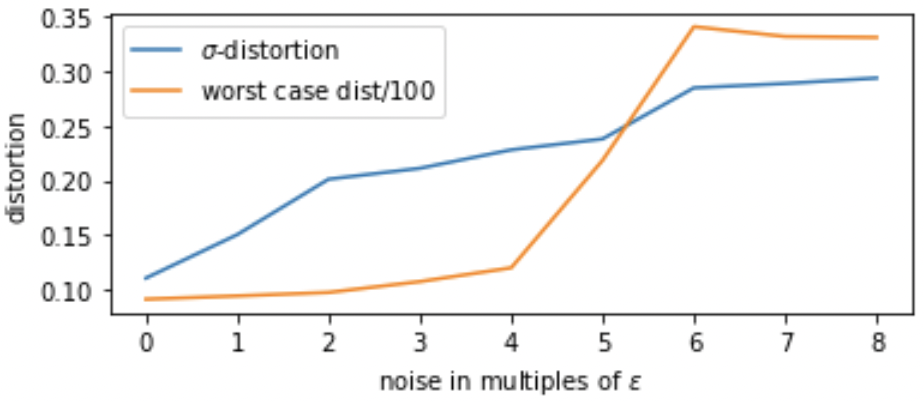}
        \caption{Local distortion of \fibred{} as a function of the added noise.}
        \label{fig:distortions}
    \end{subfigure}

    \begin{subfigure}[b]{0.19\textwidth}
        \centering
        \includegraphics[width=\linewidth]{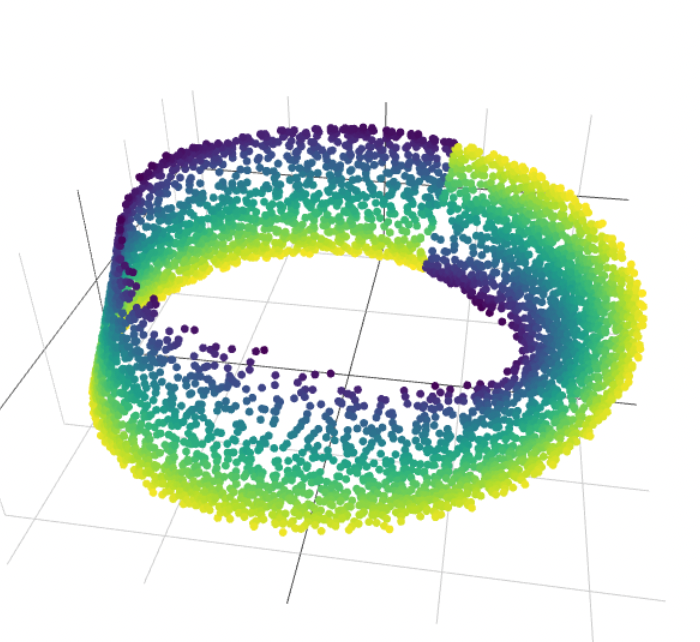}
        \caption{No noise.}
        \label{fig:noise-0}
    \end{subfigure}
    \begin{subfigure}[b]{0.19\textwidth}
        \centering
        \includegraphics[width=\linewidth]{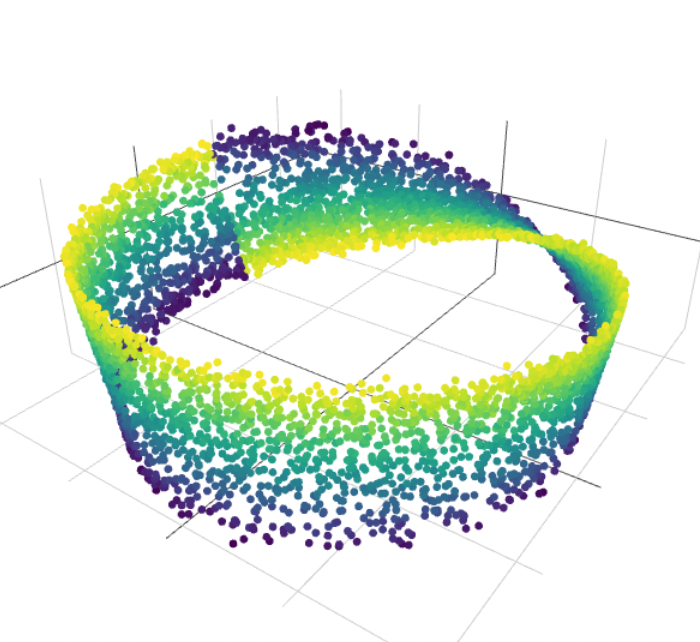}
        \caption{noise $=2\epsilon$.}
        \label{fig:noise-0}
    \end{subfigure}
    \begin{subfigure}[b]{0.19\textwidth}
        \centering
        \includegraphics[width=\linewidth]{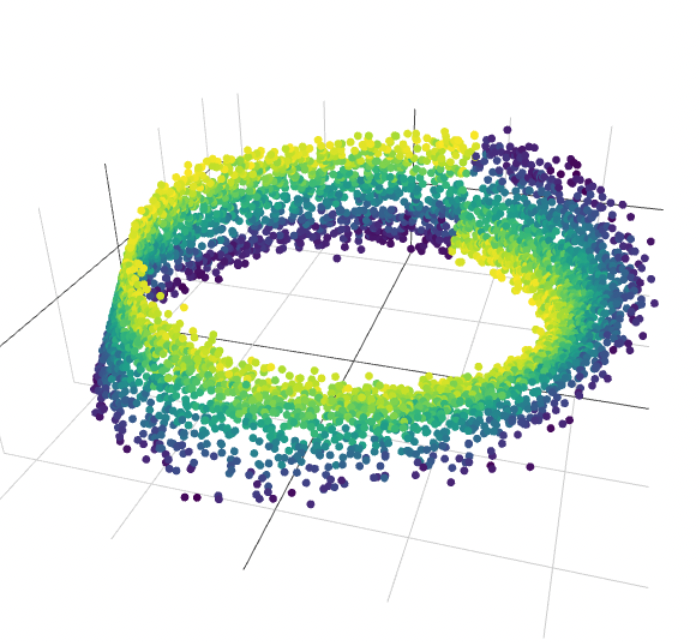}
        \caption{noise $=4\epsilon$.}
        \label{fig:noise-0}
    \end{subfigure}
    \begin{subfigure}[b]{0.19\textwidth}
        \centering
        \includegraphics[width=\linewidth]{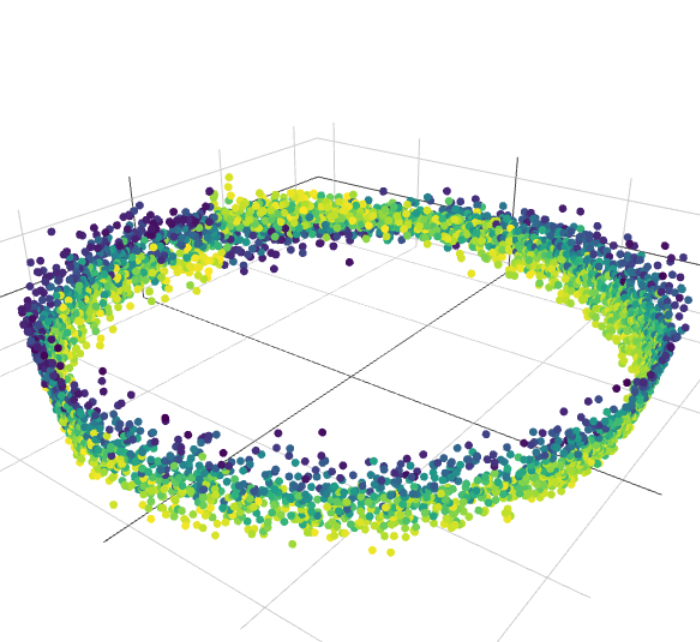}
        \caption{noise $=6\epsilon$.}
        \label{fig:noise-0}
    \end{subfigure}
     \begin{subfigure}[b]{0.19\textwidth}
        \centering
        \includegraphics[width=\linewidth]{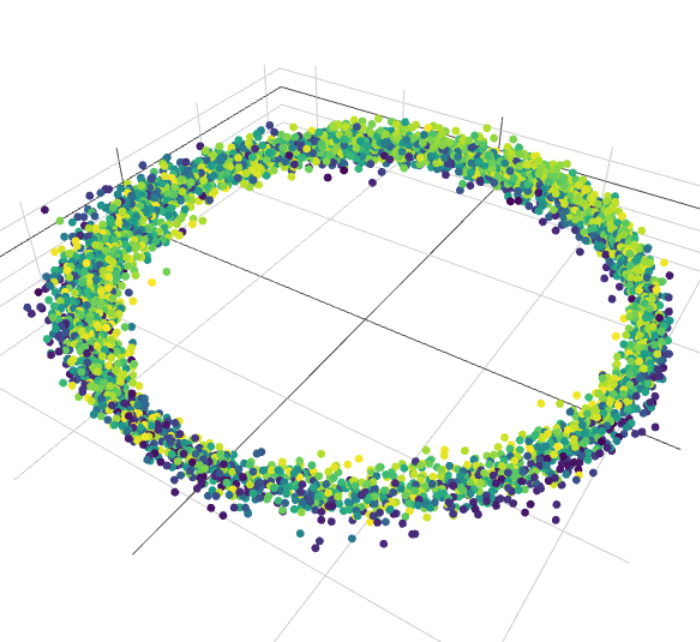}
        \caption{noise $=8\epsilon$.}
        \label{fig:noise-0}
    \end{subfigure}
      \caption{In order to isolate the behaviour of \fibred{} from the construction of the initial map, and study its robustness, we use as initial map circular coordinates on the clean data.
      We set $\epsilon$ to be the median distance from two pairs of points in the dataset.
      For each choice of noise $=2\epsilon, 4\epsilon, 6\epsilon,8\epsilon$ we sum a random number sampled uniformly from $[-\text{noise}, \text{noise}]$ to each pairwise distance between points and run \fibred{} with that distance matrix.
      We see that, even when the pairwise distances are corrupted with a large amount of noise, the non-orientability of the M\"obius band is still recovered, until it is eventually lost with noise $=8\epsilon$.
      In \cref{fig:distortions} we quantify the local distortion.
      Specifically, we take each of the cover elements obtained in the first step of \fibred{} and compute two measures of distortion (worst case distortion and $\sigma$-distortion, as described and analyzed in \cite{vankadara-luxburg}) restricting the embedding to each set in the cover and averaging over all sets.}
       \label{fig:noise-test}
\end{figure}

\begin{figure}
    \centering
    \begin{subfigure}[b]{0.28\textwidth}
        \centering
        \includegraphics[width=0.48\linewidth]{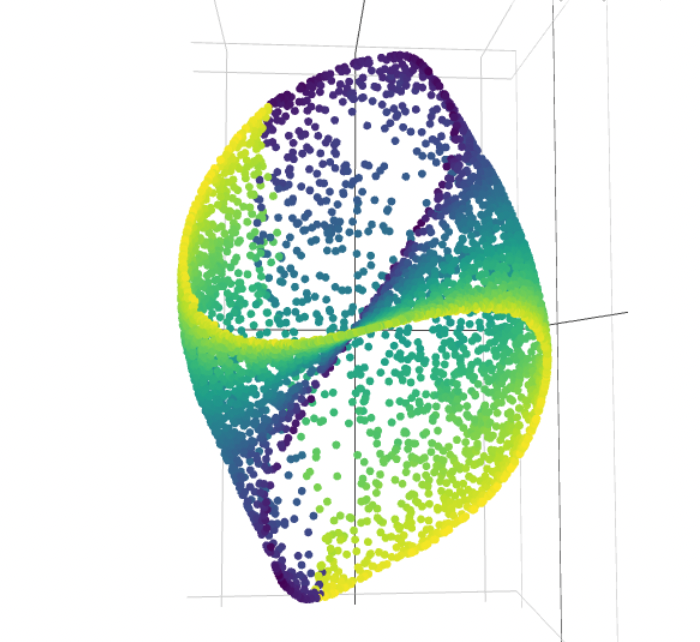}
        \includegraphics[width=0.48\linewidth]{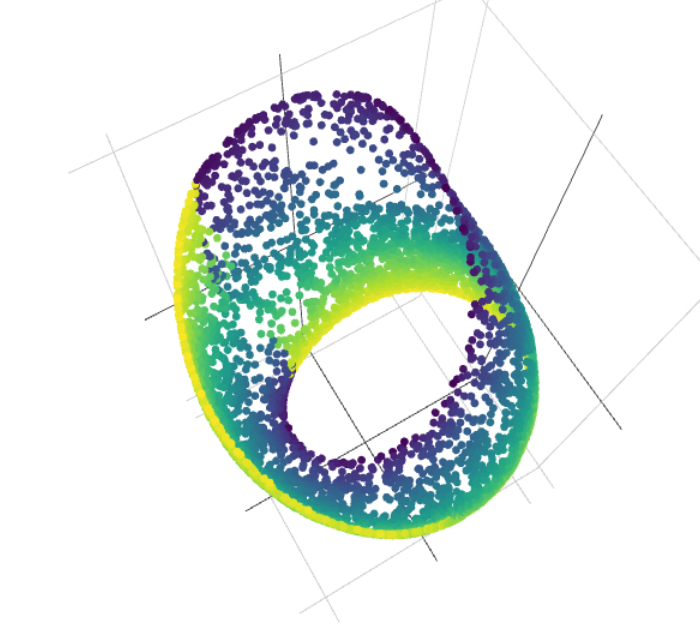}
        \caption{MDS + LE.}
        \label{fig:le-mobius}
    \end{subfigure}
     \begin{subfigure}[b]{0.2\textwidth}
        \centering
        \includegraphics[width=0.8\linewidth]{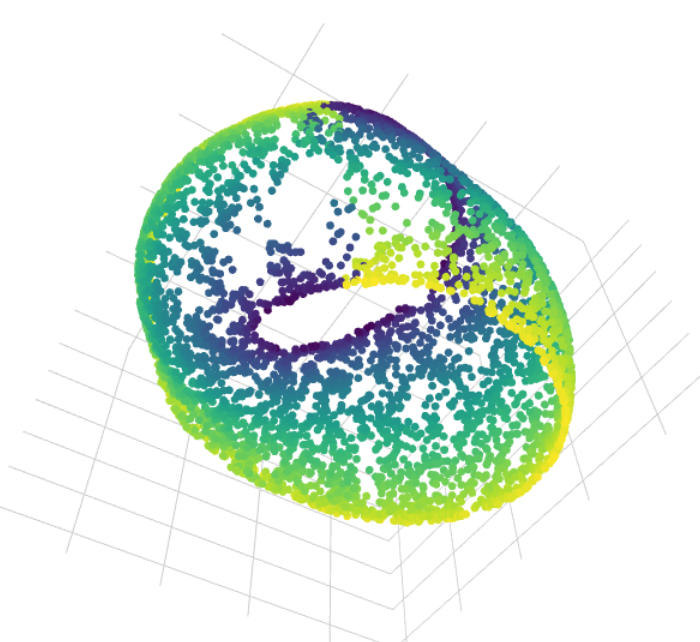}
        \caption{MDS + UMAP.}
        \label{fig:um-mobius}
    \end{subfigure}
    \begin{subfigure}[b]{0.49\textwidth}
        \centering
        {\tiny
            \begin{tabular}{|p{1cm}|p{1cm}|p{1cm}|p{1cm}|}
            \hline
                                 \rule{0pt}{2ex} & MDS + LE & MDS + UMAP & FibeRed\\ \hline
           \rule{0pt}{2ex} worst case & 0.12     & 0.14       & \textbf{0.11}    \\ \hline
           \rule{0pt}{2ex} $\sigma$-dist. & 14.74    & 9.73       & \textbf{9.06}    \\ \hline
            \end{tabular}}
        \caption{Local distortions.}
        \label{table:distortions}
    \end{subfigure}
      \caption{In order to obtain topologically faithful results with LE and UMAP on the M\"obius data, we first preprocess by running MDS with target dimension $20$.
      To quantify local distortion, we take each of the cover elements obtained in the first step of \fibred{} and compute two measures of distortion (worst case distortion and $\sigma$-distortion, as described and analyzed in \cite{vankadara-luxburg}) restricting the embedding to each set in the cover.
      We average the distortion over all sets.}
       \label{fig:other-algos-distortion}
\end{figure}

\subsection{Parameter sensitivity analysis}
\label{section:parameter-sensitivity}

We analyze the performance of \fibred{} for different choices of the cover.
In \cref{fig:cover-various-sizes} we experiment with different choices of $k$, the number of sets in the cover.
In \cref{fig:cover-random} we experiment with a not well distributed cover.

\subsection{Noise sensitivity analysis}

In \cref{fig:noise-test}, we analyze the output of \fibred{} on the M\"obis data corrupted with different amounts of noise.

\subsection{Fiber reconstruction analysis}

In \cref{fig:other-algos-distortion}, we quantify the fiberwise distortion incurred by three dimensionality reduction algorithms on the M\"obius data.

\end{document}